\documentstyle[12pt] {article}
\def\@height{height}
\def\@depth{depth}
\def\@width{width}

\newtheorem{thm}{Th\'eor\`eme}[section]
\newtheorem{lem}[thm]{Lemme}
\newtheorem{cor}[thm]{Corollaire}
\newtheorem{prop}[thm]{Proposition}

\def\g{{\gamma}}

\def\de{{\delta}}
\def\D{{\Delta}}
\def\si{{\sigma}}
\def\la{{\lambda}}
\def\La{{\Lambda}}
\def\k{{\kappa}}

\def\lp{{[ \! [}} 
\def\f{{\varphi}}

\def\th{{\theta}}
\def\Th{{\Theta}}
\def\vt{{ {\vartheta} }}

\def\e{{\varepsilon}}
\def\lb{{\langle}}\def\rb{{\rangle}}
\def\p{{\partial}}
\def\T{{\hspace {0.2cm}}}

\def\bbbr{{\rm I\!R}} 
\def\bbbn{{\rm I\!N}} 

\def\bbbc{{\mathchoice   
{\setbox0=\hbox{$\displaystyle\rm
C$}
\hbox{\hbox to0pt{\kern0.4\wd0\vrule height0.9\ht0\hss}\box0}}
{\setbox0=\hbox{$\textstyle\rm C$}\hbox{\hbox
to0pt{\kern0.4\wd0\vrule height0.9\ht0\hss}\box0}}
{\setbox0=\hbox{$\scriptstyle\rm C$}\hbox{\hbox
to0pt{\kern0.4\wd0\vrule height0.9\ht0\hss}\box0}}
{\setbox0=\hbox{$\scriptscriptstyle\rm C$}\hbox{\hbox
to0pt{\kern0.4\wd0\vrule height0.9\ht0\hss}\box0}}}}

\def\bbbq{{\mathchoice {\setbox0=\hbox{$\displaystyle\rm
Q$}
\hbox{\raise
0.15\ht0\hbox to0pt{\kern0.4\wd0\vrule height0.8\ht0\hss}\box0}}
{\setbox0=\hbox{$\textstyle\rm Q$}\hbox{\raise
0.15\ht0\hbox to0pt{\kern0.4\wd0\vrule height0.8\ht0\hss}\box0}}
{\setbox0=\hbox{$\scriptstyle\rm Q$}\hbox{\raise
0.15\ht0\hbox to0pt{\kern0.4\wd0\vrule height0.7\ht0\hss}\box0}}
{\setbox0=\hbox{$\scriptscriptstyle\rm Q$}\hbox{\raise
0.15\ht0\hbox to0pt{\kern0.4\wd0\vrule height0.7\ht0\hss}\box0}}}}

\def\bbbz{{\mathchoice {\hbox{$\textstyle\sf Z\kern-0.4em Z$}}
{\hbox{$\textstyle\sf Z\kern-0.4em Z$}}
{\hbox{$\scriptstyle\sf Z\kern-0.3em Z$}}
{\hbox{$\scriptscriptstyle\sf Z\kern-0.2em Z$}}}}

\def\R{\bbbr}
\def\N{\bbbn}
\def\Z{\bbbz}
\def\C{\bbbc}

\title{\bf {Approximations discr\`etes de la densit\'e 
 d'\'etats surfacique presque p\'eriodique} }

\date{}

\begin{document}

\maketitle

\setcounter{section}{0} 
\maketitle
\section{\'Enonc\'es des r\'esultats}  
                                                                      
\T Soit $d=d_1+d_2$ avec $d_1$, $d_2\in {\N}$.  
On identifie $x\in {\R}^d$ avec  
$(x_1,x_2)\in {\R}^{d_1}\times {\R}^{d_2}$ et  
 on s'int\'er\`esse \`a  
l'op\'erateur de Schr\"odinger sur $L^2({\R}^d)$ 
avec le potentiel continu,  
 presque p\'eriodique par rapport $x_1\in {\R}^{d_1}$ et 
d\'ecroissant suffisamment rapidement en variable 
$x_2\in {\R}^{d_2}$. Plus pr\'ecis\'ement soit $CAP({\R}^{d_1})$ 
 l'espace des fonctions continues presque p\'eriodiques 
${\R}^{d_1}\to \C$,
d\'efini comme le plus petit sous-espace ferm\'e de 
$L^{\infty}({\R}^{d_1})$ contenant toutes les fonctions 
exponentielles $\{ x_1\to {\rm e}^{i{\g}\cdot x_1} \} 
{}_{{\g}\in {\R}^{d_1}}$ et soit 
$v:\, {\R}^d\to {\R}$ la fonction v\'erifiant les hypoth\`eses 
\par \bigskip 
(H1) il existe ${\de}_0>0$ et $C>0$ tel que pour tout 
 $(x_1,x_2)\in {\R}^{d_1}\times {\R}^{d_2}$ on ait  
 $$
 |v(x_1,x_2)|\le C(1+|x_2|)^{-d_2-{\de}_0} \eqno (1.1)
 $$ 
(H2) la formule $x_2\to v(\cdot ,x_2)$ d\'efinit 
l'application continue ${\R}^{d_2}\to CAP({\R}^{d_1})$, 
o\`u $CAP({\R}^{d_1})$ est d\'efini comme au dessus avec 
la norme h\'erit\'ee de $L^{\infty}({\R}^{d_1})$.  
 \par \bigskip 

Sur l'espace de Hilbert $L^2({\R}^d)$ 
on consid\`ere l'op\'erateur auto-adjoint     
 $$
 H=-\D +V, \eqno (1.2)
 $$ 
 o\`u $\D$ est l'op\'erateur de Laplace et  
 $$
 (V\f )(x)=v(x)\f (x)\T \hbox{ pour } 
 \f \in L^2({\R}^d). \eqno (1.2')
 $$

\T Si $Z\subset {\R}^d$ alors ${\chi}_Z:\, {\R}^d\to \{ 0,\, 1\}$ 
d\'esigne la fonction caract\'eristique de $Z$ et pour
 $L,L'>0$ soit ${\chi}^{L'}_L$ l'op\'erateur d\'efini sur 
$L^2({\R}^d)$ par 
 $$
 ({\chi}^{L'}_L\f )(x)={\chi}_{[-L;\; L[^{d_1}\times 
 [-L';\; L'[^{d_2}}(x)\f (x)\T \hbox{ pour } 
 \f \in L^2({\R}^d). \eqno (1.3)
 $$  
Alors il est bien connu que 
pour toute fonction test $f\in C_0^{\infty}(\R )$ 
les op\'erateurs ${\chi}^{L'}_Lf(-\D )$ et ${\chi}^{L'}_Lf(H)$
 appartiennent \`a la classe d'op\'erateurs \`a trace 
sur ${L^2({\R}^d)}$ et on peut introduire 
 $$
 N^{L'}_L(f,H)= (2L)^{-d_1}\, 
 {\rm tr}_{L^2({\R}^d)}\, {\chi}^{L'}_L(f(H)-f(-\D )). \eqno (1.4)
 $$
Dans la Section 3 on donnera la preuve de 
\begin{thm} On suppose que le potentiel de l'op\'erateur 
de Schr\"odinger $H$ v\'erifie les hypoth\`eses {\rm (H1)} et 
{\rm (H2)}. Alors pour toute fonction $f\in C_0^{\infty}(\R )$ 
 il existe la limite th\'ermodynamique 
 $$
 N(f,H)=\lim_{L\to \infty} N^L_L(f,H). \eqno (1.5)
 $$  
\end{thm}   
 \par \smallskip  
La distribution $f\to N(f,H)$ s'appelle la densit\'e 
surfacique d'\'etats de $H$ et on donnera la preuve du fait 
qu'elle est la limite des densit\'es 
surfaciques d'\'etats des op\'erateurs aux diff\'erences finies 
$H^h$ agissant sur le reseau  
 $h{\Z}^d=\{ hn\in {\R}^d:\, n\in {\Z}^d\}$  de taille 
$h\in ]0;\; 1]$ quand $h\to 0$. \par    
\T Plus pr\'ecis\'ement soit $l^2(h{\Z}^d)$ l'espace de Hilbert  
 dont les \'el\'ements sont les applications  
 $\f :\, h{\Z}^d\to \C$ telles que 
 $$ 
 ||\f ||_h={\Bigl( \sum_{n\in {\Z}^d} |\f (hn)|^2\Bigr)}^{1/2}  
 <\infty   \eqno (1.6)
 $$
et dont le produit scalaire est donn\'e par la formule  
 $$
 {\lb \f ,\psi \rb}_h =\sum_{n\in {\Z}^d} \f (hn)
 {\overline {\psi (hn)}}. \eqno (1.6') 
 $$
Alors le laplacien discret agit sur $\f \in l^2(h{\Z}^d)$ 
selon la formule  
 $$
 ({\D}^h\f )(hn)=\sum_{j=1}^d 
 {\f (hn+he_j)-2\f (hn)+\f (hn-he_j)\over h^2} \, ,\eqno (1.7)
 $$
o\`u $e_1=(1,0,...,0),\dots ,e_d=(0,...,0,1)$ est la base 
canonique de ${\R}^d$ et soit 
 $$
 H^h={-\D}^h +V^h, \eqno (1.8)
 $$
o\`u $V^h$ est d\'efini \`a l'aide du potentiel $v$ par la formule 
 $$
 (V^h\f )(hn)=v(hn)\f (hn)\T \hbox{ pour } \f \in l^2(h{\Z}^d). 
 \eqno (1.8')
 $$ 
Par analogie au cas continu on d\'efinit 
 $$
 N^{L'}_L(f,H^h)={1\over {(2L)}^{d_1}}\! \! 
\sum_{\{ k\in {\Z}^d:\, hk\in 
 [-L;\; L[^{d_1}\times  [-L';\; L'[^{d_2}\}} \! \! 
 \lb (f(H^h)-f(-{\D}^h)){\de}_{hk},{\de}_{hk} {\rb}_h, \eqno (1.4')
 $$
o\`u ${\de}_{hk}(hn)=0$ si $n\ne k$ et ${\de}_{hk}(hk)=1$.  
Alors on a 

\begin{thm} On suppose que $H^h$ est donn\'e par $(1.8)$ avec 
$v$ v\'erifiant les hypoth\`eses {\rm (H1)} et 
{\rm (H2)}. Alors pour toute fonction $f\in C_0^{\infty}(\R )$ 
 il existe la limite th\'ermodynamique  
 $$
 N(f,H^h)=\lim_{L\to \infty} N^L_L(f,H^h). \eqno (1.9)
 $$  
De plus pour tout $\e >0$ on peut trouver $L_{\e}$, $h_{\e}>0$ 
tels que 
 $$
 \sup_{L\ge L_{\e}}\, \sup_{0<h\le h_{\e}}\,   
\left| {N(f,H^h)-N^L_L(f,H^h)}\right| \T <\e .\eqno (1.9')  
 $$
\end{thm}   
 
 \par \bigskip  
En ce qui concerne les preuves pr\'esent\'ees dans la suite, 
les r\'esultas de Th\'eor\`eme 1.1 et 1.2 concernant 
la famille des pav\'es $[-L;\; L[^d$  seront 
obtenus gr\^ace \`a l'\'etude des pav\'es 
$[-L;\; L[^{d_1}\times [-L';\; L'[^{d_2}$. 
En particulier on prouvera

\begin{thm} Soit $f\in C_0^{\infty}(\R )$ et $L'\ge 1$. Alors 
les limites   
 $$
 N^{L'}(f,H)=\lim_{L\to \infty} N^{L'}_L(f,H),\hspace{1cm}
  N^{L'}(f,H^h)=\lim_{L\to \infty} N^{L'}_L(f,H^h) \eqno (1.10)
 $$
existent et pour tout $\e >0$ on peut trouver $L_{\e}$, $h_{\e}>0$ 
tels que 
 $$
 \sup_{L\ge L_{\e}}\, \sup_{0<h\le h_{\e}}\, \left| 
 {N^{L'}(f,H^h)-N^{L'}_L(f,H^h)}\right| \T <\e .\eqno (1.10')  
 $$
De plus il existe une constante $C>0$ telle que pour tout $L'\ge 1$ 
on ait 
 $$
 |N^{L'}(f,H)-N(f,H)|\, \le CL'^{-{\de}_0}, \eqno (1.11)
 $$
 $$
 \sup_{0<h\le 1} |N^{L'}(f,H^h)-N(f,H^h)|
 \, \le CL'^{-{\de}_0}. \eqno (1.11')
 $$
\end{thm} 

 Il est \'egalement possible de retrouver $N(f,H)$ et $N(f,H^h)$ 
utilisant le proc\'ed\'e suivant 

\begin{thm} Soit $f\in C_0^{\infty}(\R )$ et $L\ge 1$.  
Alors les limites    
 $$
 N_L(f,H)=\lim_{L'\to \infty} N^{L'}_L(f,H),\hspace{7mm}
  N_L(f,H^h)=\lim_{L'\to \infty} N^{L'}_L(f,H^h) \eqno (1.12)
 $$ 
existent et on peut trouver une constante $C>0$ telle que 
pour $L,L'\ge 1$ on ait 
  $$
 |N^{L'}_L(f,H)-N_L(f,H)|\, \le CL'^{-{\de}_0},  \eqno (1.12')
 $$
 $$
 \sup_{0<h\le 1} |N^{L'}_L(f,H^h)-N_L(f,H^h)|
 \, \le CL'^{-{\de}_0}.  \eqno (1.12'')
 $$
 Si l'op\'erateur ${\chi}_L^{\infty}$ est d\'efini sur 
$L^2({\R}^d)$ par la formule  
 $$
 ({\chi}_L^{\infty}\f )(x)={\chi}_{[-L;\; L[^{d_1}\times 
 {\R}^{d_2}}(x)\f (x)\T \hbox{ pour } 
 \f \in L^2({\R}^d), \eqno (1.13)
 $$ 
alors ${\chi}_L^{\infty}(f(H)-f({-\D}))$ appartient \`a la 
classe d'op\'erateurs \`a classe sur $L^2({\R}^d)$ et on a 
les expressions
 $$
 N_L(f,H)=(2L)^{-d_1}\, {\rm tr\,} {\chi}_L^{\infty} 
 (f(H)-f({-\D})), \eqno (1.14) 
 $$
 $$
 N_L(f,H^h)={1\over {(2L)}^{d_1}}\! \! 
\sum_{\{ k\in {\Z}^d:\, hk\in [-L;\; L[^{d_1}\times {\R}^{d_2}\}}
\! \lb (f(H^h)-f({-\D}^h)){\de}_{hk},{\de}_{hk} {\rb}_h,\eqno (1.14')
 $$
o\`u la s\'erie $(1.14')$ converge absolument. De plus on a 
 $$
 N(f,H)=\lim_{L\to \infty} N_L(f,H),\hspace{1cm}
N(f,H^h)=\lim_{L\to \infty} N_L(f,H^h) \eqno (1.15)
 $$
et pour tout $\e >0$ on peut trouver $L_{\e}$, $h_{\e}>0$ 
tels que 
 $$
 \sup_{L\ge L_{\e}}\, \sup_{0<h\le h_{\e}}\, \left| 
 {N_L(H^h,f)-N(H^h,f)}\right| \T <\e .\eqno (1.15')  
 $$
\end{thm}  

Enfin dans la Section 4 on prouvera 
\begin{thm} 
Pour toute fonction $f\in C_0^{\infty}(\R )$ on a 
 $$
 N(f,H)=\lim_{h\to 0} N(f,H^h).\eqno (1.16)
 $$
\end{thm}  
Le r\'esultat du Th\'eor\`eme 1.5 permet d'obtenir
\begin{thm} (a) On d\'esigne par ${\rm supp\,} N(\cdot ,H)$ 
(respectivement ${\rm supp\,} N(\cdot ,H^h)$) le support de la 
distribution $f\to N(f,H)$ (respectivement $f\to N(f,H^h)$).
Alors 
 $$
 {\rm supp\,} N(\cdot ,H)=\{ \la \in \R :\; \lim_{h\to 0} 
 {\rm dist\,}({\rm supp\,} N(\cdot ,H^h),\la ) \, =0\} .\eqno (1.17)
 $$
(b) On d\'esigne par $\si (H)$ (respectivement 
$\si (H^h)$) le spectre de l'op\'erateur $H$ (respectivement $H^h$). 
Alors 
 $$
 \si (H)=\{ \la \in \R :\; \lim_{h\to 0} {\rm dist\,}(\si (H^h),\la )
 \, =0\} .\eqno (1.18)
 $$
(c) On a 
 $$
 \si (H^h)\cap ]-\infty ; 0[\, ={\rm supp\,} N(\cdot ,H^h)\cap 
 ]-\infty ; 0[, \eqno (1.19)
 $$
 $$
 \si (H)\cap ]-\infty ; 0[\, ={\rm supp\,} N(\cdot ,H)\cap 
 ]-\infty ; 0[. \eqno (1.19')
 $$
\end{thm}  
On peut remarquer que l'assertion (a) du Th\'eor\`eme 1.6 r\'esulte 
imm\'ediatement du Th\'eor\`eme 1.5 et l'assertion (b) r\'esulte 
des propri\'et\'es obtenues au cours de la preuve du Th\'eor\`eme 1.5. 
Ensuite (1.19) a \'et\'e d\'emontr\'e par Charour [??] et 
utilisant (1.19) 
avec (1.17), (1.18) on obtient (1.19').

\section{Id\'ees de base} 

\T Soit ${\cal B}({\cal H})$ l'alg\`ebre 
des op\'erateurs born\'es sur l'espace de Hilbert ${\cal H}$ 
et soit ${\cal B}_1({\cal H})$ l'id\'eal des op\'erateurs 
\`a trace. Si  $(e_j)_{j\in J}$ est une base orthonorm\'ee 
de l'espace de Hilbert s\'eparable ${\cal H}$ (c'est-\`a-dire 
$J$ est d\'enobrable), alors par d\'efinition 
$A\in {\cal B}_1({\cal H})\Leftrightarrow 
 \sum_{j\in J} |\lb Ae_j,\, e_j {\rb}_{\cal H}| <\infty$
et  
 $$
 {\rm tr}_{\cal H}A = \sum_{j\in J} \lb Ae_j,\, e_j {\rb}_{\cal H} 
 \T \hbox{ pour } A\in {\cal B}_1({\cal H}). \eqno (2.1)
 $$ 
Si $p\ge 1$ alors par d\'efinition 
$A\in {\cal B}_p({\cal H})\Leftrightarrow 
(A^*A)^{p/2}\in {\cal B}_1({\cal H})$ et 
 $$
||A||_{{\cal B}_p({\cal H})}={\left( { {\rm tr}_{\cal H} 
 (A^*A)^{p/2} }\right) }^{1/p}.  \eqno (2.2)
 $$ 
On abr\`ege 
 $$
 {\cal B}={\cal B}(L^2({\R}^d)), \hspace{1cm}
 {\cal B}^h={\cal B}(l^2(h{\Z}^d)),\eqno (2.3)
 $$
 $$
 {\cal B}_p={\cal B}_p(L^2({\R}^d)), \hspace{1cm}
 {\cal B}^h_p={\cal B}_p(l^2(h{\Z}^d)),\eqno (2.3')
 $$
 $$
 {\rm tr}_{L^2({\R}^d)\,}A  \, ={\rm tr\,}A, \hspace{1cm} 
 {\rm tr}_{l^2(h{\Z}^d)\,}A^h \, = {\rm tr}^h\, A^h.\eqno (2.3'') 
 $$
En introduisant l'op\'erateur ${\chi}^{h,L'}_L\in {\cal B}^h$ 
 d\'efini par 
 $$
 ({\chi}^{h,L'}_L\f )(hn)={\chi}_{[-L;\; L[^{d_1}\times 
 [-L';\; L'[^{d_2}}(hn)\f (hn)\T \hbox{ pour } 
 \f \in l^2(h{\Z}^d) \eqno (2.4)
 $$ 
et utilisant le fait que $({\de}_{hk})_{k\in {\Z}^d}$ est une base 
orthonorm\'ee de $l^2(h{\Z}^d)$ on trouve l'expression
 $$
 N^{L'}_L(f,H^h)=(2L)^{-d_1}\, {\rm tr}^h\, {\chi}^{h,L'}_L
 (f(H^h)-f({-\D}^h)). \eqno (2.5)
 $$
Dans la suite on consid\`ere le pav\'e unit\'e 
 ${\cal C}(y)=\{ x\in {\R}^d: x-y\in [0;\; 1[^d\}$ 
pour $y\in {\R}^d$ et on d\'efinit   
${\chi}_y\in {\cal B}$, \T ${\chi}^h_y\in {\cal B}^h$ \,  par 
 $$
 ({\chi}_y\f )(x)={\chi}_{{\cal C}(y)}(x)\f (x)\T \hbox{ pour } 
 \f \in L^2({\R}^d),\eqno (2.6)
 $$
 $$
 ({\chi}^h_y\f )(hn)={\chi}_{{\cal C}(y)}(hn)\f (hn)\T \hbox{ pour } 
 \f \in l^2(h{\Z}^d).\eqno (2.6')
 $$ 
Soit ${\cal K}$ un ensemble (\`a pr\'eciser plus tard) et on 
suppose que  $v_{\k}\in L^{\infty}({\R}^d)$ est r\'eelle pour 
tout $\k \in {\cal K}$. On va \'etudier les 
 op\'erateurs auto-adjoints  
 $$
 H_{\k}=-\D +V_{\k}, \hspace{1cm} H^h_{\k}=-{\D}^h +V^h_{\k}  
 \T \eqno (2.7)
 $$ 
 $$
 (V_{\k}\f )(x)=v_{\k}(x)\f (x),\hspace{9mm} 
(V^h_{\k}\f )(hn)=v_{\k}(hn)\f (hn).\eqno (2.7')
 $$ 
L'\'enonc\'e du r\'esultat cl\'e pour la suite est le suivant 
 
 \begin{prop}  Soit $f\in C_0^{\infty}(\R )$ et 
$(H_{\k})_{\k \in {\cal K}}$ 
donn\'es par $(2.7)$.  \par  
 (a)  On suppose qu'il existe $C_0>0$ telle que 
 $$
 ||V_{\k}||_{\cal B}=\sup_{x\in {\R}^d} |v_{\k}(x)|\T \le C_0
 \eqno (2.8)
 $$
pour tout $\k \in {\cal K}$. Alors il existe une constante $C>0$ 
telle que   
 $$
  ||{\chi}_yf(H_{\k})||_{{\cal B}_1} +
 ||{\chi}^h_yf(H_{\k}^h)||_{{\cal B}^h_1}\le C \eqno (2.9)
 $$  
  pour tout $\k \in {\cal K}$, $y\in {\R}^d$ et $0<h\le 1$.
\par \smallskip 
(b) Soit $\de \ge 0$. On suppose 
 $$
 M_{\de ,\k ,\k '}=\sup_{(x_1,x_2)\in {\R}^d}   (1+|x_2|)^{\de}
|v_{\k}(x_1,x_2)-v_{\k '}(x_1,x_2)|\T <\infty .\eqno (2.10)
 $$
Alors il existe $C_{\de}>0$ telle que 
 $$
 |y_2|^{\de}||{\chi}_{(y_1,y_2)}(f(H_{\k})-f(H_{\k'}))
 {\chi}_{(y_1,y_2)}||_{{\cal B}_1}\le C_{\de}M_{\de ,\k ,\k '} , 
 \eqno (2.11)
 $$
 $$
 |y_2|^{\de}||{\chi}^h_{(y_1,y_2)}(f(H^h_{\k})-f(H^h_{\k'}))   
 {\chi}^h_{(y_1,y_2)}||_{{\cal B}_1^h}\le C_{\de}M_{\de ,\k ,\k '}  
 \eqno (2.11')
 $$ 
 pour tout $\k ,\k '\in {\cal K}$, $y\in {\R}^d$ et $0<h\le 1$.
  \end{prop} 

Alors la Proposition 2.1  implique 
  
\begin{cor} (a)  Il existe une constante $C_0>0$ telle que
 $$
 \sup_{L\ge 1} |N_L^{L_1}(f,H)-N_L^{L_2}(f,H)|\le C_0
 (L_1^{-{\de}_0}+ L_2^{-{\de}_0}),\eqno (2.12)
 $$
 $$
 \sup_{L\ge 1} \sup_{0<h\le 1} |N_L^{L_2}(f,H^h)-N_L^{L_1}(f,H^h)|
 \le C_0 (L_1^{-{\de}_0}+ L_2^{-{\de}_0}).\eqno (2.12')
 $$
(b) Les Th\'eor\`emes $1.1$, $1.2$ et $1.4$ r\'esultent du 
Th\'eor\`eme $1.3$. 
\par \smallskip  
(c) Il existe une constante $C>0$ 
telle que pour tout $L$, $L'\ge 1$ on ait 
 $$
 \sup_{0<\rho \le 1} |N_{L+\rho}^{L'}(f,H)-N_L^{L'}(f,H)| 
 \le CL^{-1},\eqno (2.13)
 $$ 
 $$
 \sup_{0<\rho \le 1}\T \sup_{0<h\le 1} |N_{L+\rho}^{L'}(f,H)-
 N_L^{L'}(f,H)| \le CL^{-1}.\eqno (2.13')
 $$
\end{cor}   {\it Preuve.} (a) On suppose $L_2>L_1$. Alors 
 $$
 N_L^{L_2}(f,H^h)-N_L^{L_1}(f,H^h)=(2L)^{-d_1} {\rm tr}^h 
 ({\chi}^{h,L_2}_L-{\chi}^{h,L_1}_L)(f(H^h)-f({-\D}^h)). \eqno (2.14) 
 $$
On introduit 
 $$
 \La (L,L_1)=({\Z}^{d_1}\cap [-L;\; L[^{d_1})\times ({\Z}^{d_2}
 \setminus [ -(L_1-1);\; L_1-1[^{d_2}) \eqno (2.15)
 $$
et on remarque que 
 $$
 {\chi}^{h,L_2}_L-{\chi}^{h,L_1}_L=
 \sum_{y\in \La (L,L_1)}  \! 
 {\chi}^h_{y}({\chi}^{h,L_2}_L-{\chi}^{h,L_1}_L).\eqno (2.16)
 $$
Cependant utilisant Proposition 2.1 avec $V^h_{\k}=V^h$, 
$V^h_{\k '}=0$ on voit que $M_{\de ,\k ,\k '}<\infty$ 
avec $\de =0$ et $\de =d_2+{\de}_0$ et $(2.11')$ permet d'estimer
 $$
 \left| { {\rm tr}^h\, {\chi}^h_{(y_1,y_2)}
 ({\chi}^{h,L_2}_L-{\chi}^{h,L_1}_L)(f(H^h)-f({-\D}^h))}\right| 
 \le C(1+|y_2|)^{-d_2-{\de}_0},
 $$
donc la valeur absolue de (2.14) est major\'ee par
 $$
 \sup_{y_1\in {\Z}^{d_1}} \sum_{\{ y_2\in {\Z}^{d_2}:\; 
 |y_2|\ge L_1-1\} } C(1+|y_2|)^{-d_2-{\de}_0}\, 
 \le C_1L_1^{-{\de}_0}.
 $$ 
(b) D'abord on va v\'erifier que le Th\'eor\`eme 1.2 r\'esulte 
du Th\'eor\`eme 1.3 et du Corollaire 2.2(a). Pour obtenir 
l'assertion du Th\'eor\`eme 1.2 on doit montrer que pour tout 
$\e >0$ il est possible de trouver $L_{\e}$, $h_{\e}>0$ tels que
 $$
 L_2\ge L_1\ge L_{\e}\Rightarrow \sup_{0<h<h_{\e}} 
 |N^{L_2}_{L_2}(f,H^h)-N^{L_1}_{L_1}(f,H^h)|\T <\e .\eqno (2.17)
 $$
Soit $C_0>0$ la constante du Corollaire 2.2(a) et $L'_{\e}$ 
tel que $C_0{L'}_{\e}^{-{\de}_0} <\e /8$. Ensuite l'assertion du 
Th\'eor\`eme 1.3 permet de trouver 
$L_{\e}\ge L'_{\e}$ et $h_{\e}>0$ tels que    
 $$
 L_2\ge L_1\ge L_{\e}\Rightarrow 
 |N^{L'_{\e}}_{L_2}(f,H^h)-N^{L'_{\e}}_{L_1}(f,H^h)|\le 
 $$
 $$
 |N^{L'_{\e}}_{L_2}(f,H^h)-N^{L'_{\e}}(f,H^h)|+
 |N^{L'_{\e}}(f,H^h)-N^{L'_{\e}}_{L_1}(f,H^h)|<
 \hbox{${\e \over 2}$} \eqno (2.18)
 $$
pour $0<h<h_{\e}$. Ainsi (2.18) et $(2.12')$ impliquent   
 $$
 |N^{L_2}_{L_2}(f,H^h)-N^{L_1}_{L_1}(f,H^h)|\le 
 |N^{L'_{\e}}_{L_2}(f,H^h)-N^{L'_{\e}}_{L_1}(f,H^h)| +
 $$
 $$
 \sum_{1\le k\le 2}|N^{L_k}_{L_k}(f,H^h)-N^{L'_{\e}}_{L_k}(f,H^h)|\, 
 \le \hbox{${\e \over 2}$} +C_0(2L_{\e}^{-{\de}_0}+L_1^{-{\de}_0}+
 L_2^{-{\de}_0}) <\e  
 $$
pour $L_2\ge L_1\ge L_{\e}$, $0<h<h_{\e}$ et  
il est clair que l'assertion du Th\'eor\`eme 1.1 s'obtient  
du Th\'eor\`eme 1.3 et du Corollaire 2.2(a) de mani\`ere analogue. 
En ce qui concerne Th\'eor\`eme 1.4, il est \'evident que 
l'assertion du Corollaire 2.2(a) implique 
l'existence des limites (1.12) et les estimations 
 (1.12$'$), (1.12$''$). Ensuite pour justifier 
$(f(H)-f({-\D})){\chi}_L^{\infty}\in {\cal B}_1$ 
on remarque que la suite $((f(H)-f({-\D})){\chi}_L^{L'})_{L'\in \N}$ 
est Cauchy dans ${\cal B}_1$ et sa limite dans ${\cal B}_1$ coincide 
avec $(f(H)-f({-\D})){\chi}_L^{\infty}$ car 
 $\lim_{L'\to \infty} ||{\chi}_L^{\infty}\f -{\chi}_L^{L'}\f ||\, =0$ 
 pour tout $\f \in L^2({\R}^d)$. 
De la m\^eme mani\`ere on trouve que 
$((f(H^h)-f({-\D}^h)){\chi}_L^{L'})_{L'\in \N}$ converge dans 
${\cal B}^h_1$ vers $(f(H^h)-f({-\D}^h)){\chi}_L^{h,\infty}$, o\`u 
 $$
 ({\chi}_L^{h,\infty}\f )(hn)={\chi}_{[-L;\; L[^{d_1}\times 
 {\R}^{d_2}}(hn)\f (hn)\T \hbox{ pour } 
 \f \in l^2(h{\Z}^d). \eqno (2.19)
 $$ 
Ainsi on a
  $$
 N_L(f,H^h)=(2L)^{-d_1}\, {\rm tr\,} {\chi}_L^{h,\infty} 
 (f(H^h)-f({-\D}^h))  \eqno (2.20) 
 $$
et par cons\'equent la s\'erie $(2.14')$ converge absolument.
Finalement pour tout $\e >0$ il existe $L'_{\e}>0$ tel que 
 $$
 L\le L'_{\e}\Rightarrow \sup_{0<h\le 1} \sup_{L\ge 1} |N_L(f,H^h)
 -N_L^L(f,H^h)|\le C_0L'^{-{\de}_0}<\e /2,
 $$
donc $(1.15'')$ r\'esulte de $(1.9')$.  
\par \smallskip  
(c)   
Soit ${\La}'(L)={\Z}^{d_1}\cap ([-(L+2);\; L+2]^{d_1}\setminus 
 [-(L-1);\; L-1]^{d_1})$. Alors 
 $$
 {\chi}^{h,L'}_{L+\rho}-{\chi}^{h,L'}_L=
 \sum_{y\in {\La}'(L)\times {\Z}^{d_2}} \! \! \!
 {\chi}^h_{y}({\chi}^{h,L_2}_L-{\chi}^{h,L_1}_L) \eqno (2.21)
 $$
et compte tenu du fait que card$\, {\La}'(L)\, \le C_0L^{d-1}$ on a
peut estimer le membre gauche de (1.13$'$) par
 $$
 C_1L^{-1} \sup_{y_1\in {\Z}^{d_1}} \sum_{y_2\in {\Z}^{d_2}} 
 ||{\chi}_{(y_1,y_2)}(f(H^h)-f({-\D}^h))
 {\chi}_{(y_1,y_2)}||_{{\cal B}_1} 
 $$
 $$
 \le C_2L^{-1}\sum_{y_2\in {\Z}^{d_2}} 
 (1+|y_2|)^{-d_2-{\de}_0}\, \le CL^{-1}. \eqno \triangle 
 $$

\par \bigskip 

\begin{cor} Pour prouver {\rm Th\'eor\`eme 1.5} il suffit 
de montrer que pour toute fonction 
$\th \in C_0^{\infty}({\R}^d)$ on a 
 $$
 \lim_{h\to 0} {\rm tr}^h\, {\Th}^h(f(H^h)-f({-\D}^h))\, =
\, {\rm tr}\,  {\Th}(f(H)-f(-\D )),\eqno (2.22)
 $$  
o\`u ${\Th}$ et ${\Th}^h$ sont les op\'erateurs de multiplication
 $$
 ({\Th}\f )(x)={\th}(x)\f (x)\, \hbox{ pour } 
 \f \in L^2({\R}^d),\eqno (2.23)
 $$ 
 $$
 ({\Th}^h\f )(hn)={\th}(hn)\f (hn)\, \hbox{ pour } 
 \f \in l^2(h{\Z}^d).\eqno (2.23')
 $$ 
\end{cor} 
{\it Preuve.} L'assertion du Th\'eor\`eme 1.4 sera d\'emontr\'ee 
si on montre que pour tout $\e >0$ il est possible 
de trouver $L_{\e}$, $h_{\e }>0$ tels que
 $$
 L\ge L_{\e }\Rightarrow \sup_{0<h<h_{\e }} 
 |N^L_L(f,H^h)-N^L_L(f,H^h)|\T <\e .\eqno (2.24)
 $$
Soit $C_0>0$ la constante du Corollaire 2.2(a) et $L_{\e}$ 
tel que $C_0L_{\e}^{-{\de}_0}<\e /8$. Ensuite pour $L>0$ on peut 
choisir ${\th}_L\in C_0^{\infty}({\R}^d)$ telle que 
$0\le {\th}_L\le 1$, ${\th}_L=1$ sur $[-L;\; L[^d$ et 
${\rm supp\,} {\th}_L\subset [-L-1;\; L+1[^d$. On introduit 
les op\'erateurs ${\Th}_L$ et ${\Th}_L^h$ d\'efinis par
 $$
 ({\Th}_L\f )(x)={\th}_L(x)\f (x)\, \hbox{ pour } 
 \f \in L^2({\R}^d),
 $$ 
 $$
 ({\Th}_L^h\f )(hn)={\th}_L(hn)\f (hn)\, \hbox{ pour } 
 \f \in l^2(h{\Z}^d)
 $$  
et on montre qu'il est possible de choisir $L_{\e }$ tel que
 $$
 L\ge L_{\e }\Rightarrow \sup_{0<h<h_{\e }} 
 |N^L_L(f,H^h)-{\rm tr}^h\, {\Th}_L^h(f(H^h)-f({-\D}^h))|\T <\e /4,
 \eqno (2.25)
 $$
 $$
 L\ge L_{\e }\Rightarrow |N^L_L(f,H)-{\rm tr}\, 
 {\Th}_L(f(H)-f(-\D ))|\T <\e /4.\eqno (2.25')
 $$
En effet pour obtenir (2.24) on remarque que 
 $$
 {\Th}_L^h-{\chi}^{h,L}_L=
\sum_{y\in \La (L,L)\cup (\La '(L)\times {\Z}^{d_2})} 
 {\chi}_y^h ({\Th}_L^h-{\chi}^{h,L}_L)
 $$
o\`u $\La (L,L)$ et $\La '(L)$ sont comme dans la preuve du 
Corollaire 2.2. Alors de la m\^eme mani\`ere on trouve 
 $$
 \sum_{y\in \La '(L)\times {\Z}^{d_2}} |{\rm tr}^h\, ({\Th}_L^h-
 {\chi}^{h,L}_L){\chi}_y^h(f(H^h)-f({-\D}^h))|\, \le C_0L^{-{\de}_0},
 $$
 $$
 \sum_{y\in \La (L,L)} |{\rm tr}^h\, ({\Th}_L^h-{\chi}^{h,L}_L)
 {\chi}_y^h(f(H^h)-f({-\D}^h))|\, \le C_1L^{-1}  
 $$
et (2.25) est assur\'e si 
$C_0L_{\e }^{-{\de}_0}+C_1L_{\e }^{-1}<\e /4$. 
De mani\`ere analogue on obtient (2.25$'$) et pour terminer la preuve 
de (2.24) il suffit de remarquer que 
(2.22) assure l'existence de $h_{\e}>0$ tel que
 $$
 \sup_{0<h<h_{\e }} \left| {{\rm tr}^h\, {\Th}_L^h(f(H^h)-f({-\D}^h)
 )\, -\, {\rm tr}\,  {\Th}_L(f(H)-f(-\D )) }\right| <\e /2.
 \eqno \triangle 
 $$
\par \bigskip 
\begin{lem} Soit $\e >0$. Alors on peut trouver $N(\e )\in \N$, 
${\g}_{\e ,k}\in {\R}^{d_2}$, 
$v_{\e ,k}\in C_0^{\infty }({\R}^{d_2})$ 
pour $k=1,\dots ,N(\e )$ tels que 
 $$
 \left| {v(x_1,x_2)-\sum_{k=1}^{N(\e )} v_{\e ,k}(x_2)
 e^{ix_1{\g}_{\e ,k}} }\right| <\e . 
 $$
o\`u $e^h_{{\g}_{\e ,k}}\in {\cal B}(l^2(h{\Z}^{d_1}))$, 
$V^h_{\e ,k}\in {\cal B}(l^2(h{\Z}^{d_2}))$ sont donn\'es par 
 $$
 (e^h_{{\g}_{\e ,k}}{\f}_1)(hn_1)={\rm e}^{ihn_1{\g}_{\e ,k}}{\f}_1   (hn_1)\T \hbox{ pour } \T {\f}_1\in l^2(h{\Z}^{d_1}),
 $$ 
 $$
 (V^h_{\e ,k}{\f}_2)(hn_2)=v_{\e ,k}(hn_2){\f}_2(hn_2)
 \T \hbox{ pour } \T {\f}_2\in h^2(h{\Z}^{d_2}).
 $$
\end{lem} 
{\it Preuve.} Soit $\e >0$ et 
$w_{\e}(x_1,x_2)=v(x_1,x_2){\th}_{\e}(x_2)$ o\`u 
${\th}_{\e}\in C_0^{\infty}({\R}^{d_2})$ est choisie telle que 
$0\le {\th}_{\e}\le 1$ et $|v(x)-w_{\e}(x)|<\e /2$ 
pour tout $x\in {\R}^d$. Ensuite on remarque que pour tout 
$\e '>0$ il existe $\de '>0$ tel que 
 $$
 |x_2-x'_2|<\de '\, \Rightarrow \, \sup_{x_1\in {\R}^{d_1}} 
 |w_{\e}(x_1,x_2)-w_{\e}(x_1,x'_2)|\, <\e '
 $$
et utilisant p. ex. les formules de polyn\^omes de Bernstein 
on peut trouver les coefficients $c_{N,y_2,\nu ,\e}\in \R$ 
tels que pour tout $x_1\in {\R}^{d_1}$, 
$x_2\in {\rm supp\,} {\th}_{\e}$ on ait
 $$
 \left| {v(x_1,x_2)-
 \sum_{ {}^{y_2\in {\Z}^{d_2}\cap 
 {\rm supp\,}{\th}_{\e}}_{\{ \nu \in {\Z}^{d_2}:\; |\nu |\le 
 2d_1N(\e )\} } } 
 v(x_1,y_2/N(\e ))c_{N(\e ),y_2,\nu ,\e}x_2^{\nu} }\right| \, 
 <{\e \over 4}  
 $$
si $N(\e )\in \N$ est choisi suffisamment grand. \par 
Pour terminer la preuve on remarque que 
$v(\cdot ,N(\e )^{-1}y_2)\in CAP({\R}^{d_1})$ permet 
de trouver $N'(\e )$ et $c_{\e ,k,y_2}\in \C$, 
 ${\g}_{\e ,k,y_2}\in {\R}^{d_1}$ 
pour $k=1,\dots ,N'(\e )$ tels que 
 $$
 \left| {w_{\e}(x_1,x_2)- 
 \sum_{{}^{y_2\in {\Z}^{d_2}\cap 
 {\rm supp\,}{\th}_{\e}}_{1\le k\le N'(\e )}} c_{\e ,k,y_2}
 {\rm e}^{ix_1{\g}_{\e ,k,y_2}} v_{\e ,k,y_2}(x_2) }\right| \, 
 <{\e \over 2}  
 $$
avec 
 $$
 v_{\e ,k,y_2}(x_2)=
 \sum_{\{ \nu \in {\Z}^{d_2}:\; |\nu |\le 2d_1N(\e )\} } 
 c_{N(\e ),y_2,\nu ,\e}x_2^{\nu}{\th}_{\e}(x_2).\eqno \triangle
 $$

\section{Preuve de Th\'eor\`emes 1.1-1.4} 

\T La preuve est bas\'ee  sur l'\'etude de la 
famille d'op\'erateurs $(H_z)_{z\in {\R}^{d_1}}$
d\'efinis sur $L^2({\R}^d)$ par la formule
 $$
 H_z=-\D +V_z, \eqno (3.1)
 $$
 $$
 (V_z\f )(x)=v(x_1+z,x_2)\f (x)\T \hbox{ pour } 
 \f \in L^2({\R}^d). \eqno (3.1')
 $$ 
\par \smallskip 
On commence par l'\'enonc\'e du r\'esultat cl\'e de cette section : 			 
  
\begin{prop} Soit $f\in C_0^{\infty}(\R )$, $L'\in \N$ et  
$y_1\in {\Z}^{d_1}$. On pose
 $$
 u^{L'}_{y_1}(z)=\sum_{y_2\in {\lp -L';\; L'\lp}^{d_2} } 
 {\rm tr\,} {\chi}_{(y_1,y_2)} (f(H_z)-f(-\D )),\eqno (3.2)
 $$
o\`u \T $\lp -L';\; L'\lp =\Z \cap [-L'; L'[$. \T 
Alors $u^{L_0}_{y_1}\in CAP({\R}^{d_1})$.
\end{prop} 
{\underline {\it Preuve du fait que Proposition $3.1$ implique 
Th\'eor\`eme $1.1$} } 
\par \smallskip 
Soit $T_{(z,0)}$ l'op\'erateur de la translation, 
$(T_{(z,0)}\f )(x)=\f (x_1-z,x_2)$. Alors 
 $$
 H=T_{(z,0)}H_zT_{(-z,0)},\hspace{1cm} {\chi}_{(y_1,y_2)}=
T_{(y_1,0)} {\chi}_{(y_1,y_2)}T_{(-y_1,0)},\eqno (3.3)
 $$
 $$
 {\rm tr\,} T_{(-y_1,0)}{\chi}_{(0,y_2)} f(H_{y_1}) T_{(y_1,0)}\, 
={\rm tr\,} {\chi}_{(y_1,y_2)} f(H)\eqno (3.4)
 $$
et il est clair que 
 $$
 u^{L'}_{y_1}(0)=u^{L'}_0(y_1).\eqno (3.5)
 $$
Ainsi notant par $[L]$ la partie enti\`ere de $L$ on peut \'ecrire 
 $$
  (2[L])^{-d_1}\sum_{y_1\in {\lp -L; L\lp}^{d_1}} u^{L'}_0(y_1)\, 
 =N^{L'}_{[L]}(f,H), \eqno (3.6)
 $$
et compte tenu du Corolaire 2.2 il suffit de montrer que pour tout 
$\e >0$ il existe $L_{\e}>0$ tel que 
 $$
 L_2\ge L_1\ge L_{\e}\Rightarrow 
 |N^{L'}_{L_1}(f,H)-N^{L'}_{L_2}(f,H)|<\e .\eqno (3.7)
 $$
Soit $C_1$ la constante de l'assertion du Corollaire 2.2(c) et 
soit $L_{\e}\ge \e /(4C_1)$. Alors pour obtenir (3.7) 
il suffit de montrer  
 $$
 L_2\ge L_1\ge L_{\e}\Rightarrow 
 |N^{L'}_{[L_1]}(f,H)-N^{L'}_{[L_2]}(f,H)|<\e /2.\eqno (3.8)
 $$
Par d\'efinition de $CAP({\R}^{d_1})$ on peut trouver 
$N(\e )\in \N$, $c_{\e ,k}\in \C$, ${\g}_{\e ,k}\in {\R}^{d_1}$ 
tels que l'on ait 
 $$
 |u^{L'}_0(z)-u^{L'}_{\e}(z)|<\e /8 \eqno (3.9)
 $$ 
avec 
 $$
 u^{L'}_{\e}(z)=\sum_{k=0}^{N(\e )} 
 c_{\e ,k}{\rm e}^{iz{\g}_{\e ,k}}. \eqno (3.10)
 $$
Alors il suffit de montrer que pour tout $\e >0$, la fonction   
 $$ 
 N^{L'}_{\e ,[L]}(f,H)=(2[L])^{-d_1}
 \sum_{y_1\in {\lp -L; L\lp }^{d_1}} u^{L'}_{\e}(y_1)\eqno (3.11)
 $$
poss\`ede une limite quand $L\to \infty$. En effet, 
la condition de Cauchy pour $L\to N^{L'}_{\e ,[L]}(f,H)$ assure 
l'existence de $L_{\e}>0$ tel que
 $$
 L_2\ge L_1\ge L_{\e}\Rightarrow |N^{L'}_{\e ,[L_2]}(f,H) 
 -N^{L'}_{\e ,[L_1]}(f,H)|<\e /4 \eqno (3.12)
 $$
et (3.9) permet d'estimer    
 $$
 |N^{L'}_{[L]}(f,H)-N^{L'}_{\e ,[L]}(f,H)|\le \e /8 \eqno (3.13)
 $$ 
pour $L=L_1$ et $L=L_2$, donc l'in\'egalit\'e triangulaire 
implique (3.8). \par 
  Il reste \`a justifier que pour tout $\g \in {\R}^{d_1}$ 
la fonction 
 $$
 L\to N_{[L]}(\g )= (2[L])^{-d_1} 
\sum_{y_1\in {\lp -L;\; L\lp}^{d_1}} {\rm e}^{iy_1\g}\eqno (3.14)
 $$
poss\`ede une limite quand $L\to \infty$. 
Cependant dans le cas $\g \in 2\pi {\Z}^{d_1}$, 
l'assertion est \'evidente car 
 $$
 |1-N_{[L]}(\g )|=|1-(2[L])^{-d_1}{\rm card\,} 
 {\lp -L;\; L\lp}^{d_1}|
 \le C/L\to 0\T \hbox{ quand } L\to \infty .
 $$ 
Il reste \`a \'etudier le cas 
$\g =({\g}{(1)},...,{\g}{(d_1)})\notin 2\pi {\Z}^{d_1}$.  
Si $j_0\in \{ 1,...,d\}$ est tel que 
${\g}{(j_0)}\notin 2\pi {\Z}$, alors utilisant 
 $$
 \left| { \sum_{\nu = -L}^{L-1}  
 {\rm e}^{i\nu {\g}{(j_0)}} }\right| \le 
 {2\over |{\rm e}^{i{\g}{(j_0)}}-1| } \eqno (3.15)
 $$ 
et $|N_{[L]}(\g )|=\Pi_{j=1}^{d_1} \left| {{1\over 2[L]}
 \sum_{\nu = -L}^{L-1}  {\rm e}^{i\nu {\g}{(j)}}  }\right| 
 \le 2L^{-1}|{\rm e}^{i{\g}{(j_0)}}-1|^{-1}\to 0$ 
quand $L\to \infty$.   $\triangle$
\par \bigskip 
 
\T Avant de commencer la preuve de Proposition 3.1 on va introduire  
des notations auxiliaires. Si ${\cal A}$ est un espace de Banach, 
alors $C_{\rm pp}({\R}^{d_1};\; {\cal A})$ d\'esigne 
l'ensemble des applications $\Phi :\, {\R}^{d_1}\to {\cal A}$
telles que  pour tout $\e >0$ on peut 
trouver $N(\e )\in \N$, ${\g}_{\e ,k}\in {\R}^{d_1}$, 
$A_{\e ,k}\in {\cal A}$ $(k=1,...,N(\e ))$ tels que 
 $$
 \sup_{z\in {\R}^{d_1}}||\Phi (z) -{\sum}_{k=1}^{N(\e )} A_{\e ,k}\, 
{\rm e}^{iz\cdot {\g}_{\e ,k}} ||_{\cal A}\, <\e .\eqno (3.16)
 $$ 
On remarque que $C_{\rm pp}({\R}^{d_1};\; {\cal A})$ est un espace 
de Banach avec la norme  
 $$
 ||\Phi ||_{\infty}=\sup_{z\in {\R}^{d_1}} ||\Phi (z)||_{\cal A}.
 \eqno (3.17)
 $$ 
\par \smallskip   
\begin{lem} Si $V_z$ est donn\'e par $(3.1')$, alors la fonction 
$z\to V_z$ appartient \`a 
$C_{\rm pp}({\R}^{d_1};\; {\cal B})$. 
\end{lem} 
{\it Preuve.} Soit $\e >0$. Alors on a 
$||V-V_{0,\e}||_{\cal B}<\e$ avec 
 $V^{0,\e}=\sum_{k=0}^{N(\e )} V_{\e ,k},$
o\`u $V_{\e ,k}$ d\'esigne l'op\'erateur de multiplication par 
$v_{\e ,k}(x)={\rm e}^{i{\g}_{\e ,k}x_1}{\tilde v}_{\e ,k}(x_2)$ 
comme dans la Section 2. 
Alors 
 $$
 V_{z,\e}=T_{(z,0)}^{-1}V_{0,\e}T_{(z,0)}=\sum_{k=0}^{N(\e )} 
 V_{\e ,k}{\rm e}^{iz{\g}_{\e ,k}} \eqno (3.17)
 $$
et on a \T 
$||V_z-V_{z,\e}||_{\cal B}=||T_{(z,0)}^{-1} (V-V_{0,\e})T_{(z,0)}
||_{\cal B}<\e$. 
 $\triangle$
\par \bigskip

\begin{lem} Soit $p\ge 1$. Alors 
$C_{\rm pp}({\R}^{d_1};\; {\cal B}_p)$ est un id\'eal 
bilat\'eral de l'alg\`ebre de Banach 
 $C_{\rm pp}({\R}^{d_1};\; {\cal B})$. Si   
${\Phi}_j\in C_{\rm pp}({\R}^{d_1};\; {\cal B}_{p_j})$ 
pour  $j=1$, $2$, alors on a ${\Phi}_1{\Phi}_2\in 
C_{\rm pp}({\R}^{d_1};\; {\cal B}_p)$ avec  
$p=p_1p_2/(p_1+p_2)$.   
\end{lem} 
{\it Preuve.} Si ${\Phi}_j\in C_{\rm pp}({\R}^{d_1};\; 
{\cal B}_{p_j})$ et $\e >0$, alors on peut trouver $N(\e )$ et 
$A_{\e ,k,j}\in {\cal B}_{p_j}$, 
${\g}_{\e ,k,j}\in {\R}^{d_1}$ ($k=1,...,N{\e}$) tels que
 $$
 {\Phi}_{\e ,j}(z)={\sum}_{k=1}^{N(\e )} 
 A_{\e ,k,j}\, {\rm e}^{iz\cdot {\g}_{\e ,k,j}} \eqno (3.18)
 $$ 
v\'erifie $||{\Phi}_j(z)-{\Phi}_{\e ,j}(z)||_{{\cal B}_{p_j}}<
\e (1+||{\Phi}_1||_{\infty}+||{\Phi}_2||_{\infty})^{-1}$. 
Puisque   
 $$
 \hbox{${1\over p}={1\over p_1}+{1\over p_2}$} \, \Longrightarrow 
 ||A_{\e ,k,1}A_{\e ,k',2}||_{{\cal B}_p}\le ||A_{\e ,k,1}
 ||_{{\cal B}_{p_1}}||A_{\e ,k',2}||_{{\cal B}_{p_2}}, \eqno (3.19)
 $$ 
on peut trouver $N_{\e}$ et $A_{\e ,k}\in {\cal B}_p$, 
${\g}_{\e ,k}\in {\R}^{d_1}$ ($k=1,...,N{\e}^2$) tels que 
 $$
 {\Phi}_{\e ,1}(z){\Phi}_{\e ,2}(z)={\sum}_{k=1}^{N{\e}^2} 
 A_{\e ,k}\, {\rm e}^{iz\cdot {\g}_{\e ,k}} \eqno (3.18')
 $$ 
et on peut estimer $||{\Phi}_1(z){\Phi}_2(z)-{\Phi}_{\e ,1}(z)
{\Phi}_{\e ,2}(z)||_{{\cal B}_p}$ par 
 $$
  ||({\Phi}_1(z)-{\Phi}_{\e ,1}(z)){\Phi}_2(z)
 ||_{{\cal B}_p} +||{\Phi}_{\e ,1}(z)({\Phi}_2(z)-
 {\Phi}_{\e ,2}(z)) ||_{{\cal B}_p} \le
 $$
 $$
  ||{\Phi}_1(z)-{\Phi}_{\e ,1}(z)||_{{\cal B}_{p_1}} 
 ||{\Phi}_2(z)||_{{\cal B}_{p_2}}+||{\Phi}_{\e ,1}(z)
 ||_{{\cal B}_{p_1}} ||{\Phi}_2(z)-{\Phi}_{\e ,2}(z) 
 ||_{{\cal B}_{p_2}} <\e .
 $$ 
\par \smallskip
On en d\'eduit ${\Phi}_1{\Phi}_2\in C_{\rm pp}({\R}^{d_1};\; 
{\cal B}_p)$ et il est \'evident qu'en rempla{\c {c}}ant 
${\cal B}_{p_j}$ par ${\cal B}$ dans ce 
raisonnement on trouve que 
$C_{\rm pp}({\R}^{d_1};\; {\cal B})$ est une alg\`ebre. 
Enfin utilisant le raisonnement analogue avec
  $$
 ||A_{\e ,k,1}A_{\e ,k',2}||_{{\cal B}_p}\le 
||A_{\e ,k,1}||_{{\cal B}}
 ||A_{\e ,k',2}||_{{\cal B}_p} \eqno (3.20)
 $$   
on trouve que 
$C_{\rm pp}({\R}^{d_1};\; {\cal B}_p)$ est un id\'eal 
\`a gauche et pareillement 
  $$
 ||A_{\e ,k,1} A_{\e ,k',2}||_{{\cal B}_p}\le 
 ||A_{\e ,k,1}||_{{\cal B}_p}
 ||A_{\e .k',2} ||_{{\cal B}}  \eqno (3.20')
 $$  
permet de trouver que 
$C_{\rm pp}({\R}^{d_1};\; {\cal B}_p)$ est un id\'eal 
\`a droite. $\triangle$
\par \bigskip 
 
On utilisera le r\'esultat suivant 
 
\begin{prop} Soit ${\la}_0\ge 1+2||V||_{\cal B}$ et pour 
$m\in {\N}^*$ on pose  
 $$ 
 R_z^m=(-\D +V_z+{\la}_0I)^{-m}, \hspace{8mm} 
(R_z^h)^m=(-{\D}^h +V^h_z+{\la}_0I)^{-m}.\eqno (3.21)
 $$
Si $p\ge \max \{ m,\, 1+{d\over 2} \}$ et $N\in \N$, alors on peut 
trouver des constantes $C_{N,m}>0$ telles que l'on ait
 $$
 || {\chi}_y R_z^m||_{{\cal B}_{p/m}}+||{\chi}^h_y(R_z^h)^m
 ||_{{\cal B}^h_{p/m}} \le C_{0,m},\eqno (3.22)
 $$ 
 $$
 || {\chi}_y R_z^m{\chi}_{y'}||_{{\cal B}_{p/m}}+
 || {\chi}^h_y (R_z^h)^m{\chi}^h_{y'}||_{{\cal B}^h_{p/m}}
 \le C_{N,m} (1+|y-y'|)^{-N} \eqno (3.22')
 $$ 
pour tout $y,y'\in {\Z}^d$ et $0<h\le 1$. 
\end{prop} 
La preuve de la Proposition 3.4 sera d\'etaill\'ee dans 
le Chapitre 3. 
\par \smallskip   

{\underline {\it Preuve de la Proposition $3.1$.} } 
\par \smallskip 
On fixe $m_0\in \N$ tel que $m_0\ge 1+{d\over 2}$. 
Soit $\e >0$. Alors le th\'eor\`eme de Weierstrass 
permet de trouver un polyn\^ome d'une variable r\'eelle, 
$g_{\e}(s)=\sum_{k=0}^{N(\e )} c_{\e ,k}s^k$, tel que 
 $$
 \sup_{\la \ge {\la}_0/2} \left| {(\la +{\la}_0)^{m_0}f(\la )-
 g_{\e}\left( {(\la +{\la}_0)^{-1}}\right) } \right| 
 <\e /C_{0,m_0},\eqno (3.23)
 $$ 
o\`u $C_{0,m_0}$ est la constante de la formule (3.22).
En posant $f_{\e}(\la )=(\la +{\la}_0)^{-m_0}
g_{\e}\left( {(\la +{\la}_0)^{-1}}\right)$ on obtient 
 $$
 \la \ge {\la}_0/2\, \Rightarrow \pm \, (f(\la )-f_{\e}(\la ))
 \le \e (\la +{\la}_0)^{-m_0}/C_{0,m_0},
 $$
donc 
 $$
 \pm \, {\rm tr\,} {\chi}_y(f(H_z)-f_{\e}(H_z)){\chi}_y   
 \le {\e \over C_{0,m_0}} 
 {\rm tr\,} {\chi}_yR_z^{m_0} {\chi}_y   <\e .\eqno (3.24)
 $$ 
Ainsi il suffit de prouver que pour tout $\e >0$ la fonction 
 $$
 z\to {\rm tr\,} {\chi}_yf_{\e}(H_z)\, =\sum_{k=0}^{N(\e )} 
 c_{\e ,k} {\rm tr\,} {\chi}_yR_z^{m_0+k}
 $$
appartient \`a $CAP({\R}^{d_1})$. Pour cela 
on va montrer que les fonctions 
 $z\to {\Phi}_{y,m}(z)={\chi}_yR_z^m$ appartiennent \`a 
$C_{\rm pp}({\R}^{d_1},{\cal B}_1)$ pour $m\ge m_0$. 
Plus pr\'ecis\'ement on va \'etablir 
 $$
 p\ge \max \{ m,\, 1+\hbox{${d\over 2}$}\} \Rightarrow {\Phi}_{y,m}
 \in C_{\rm pp}({\R}^{d_1},{\cal B}_{p/m}) \eqno (3.25(m))
 $$  
par r\'ecurrence par rapport \`a $m\in {\N}^*$. \par   
 On commence par $m=1$. On pose $R=(-\D +{\la}_0I)^{-1}$ et  
 $$
 {\Phi}_{y,1,N}(z)={\chi}_yR\sum_{k=0}^N (-V_zR)^k.\eqno (3.26)
 \T 
 $$ 
 Puisque $||V_zR||_{\cal B}\le 1/2$ on peut \'ecrire 
 $$
 {\Phi}_{y,1}(z)={\chi}_yR(I+V_zR)^{-1}={\chi}_yR
 \sum_{k=0}^{\infty} (-V_zR)^k\eqno (3.27)
 $$ 
et on trouve 
 $$
 ||{\Phi}_{y,1}(z)-{\Phi}_{y,1,N}(z)||_{{\cal B}_p}\le 
 ||{\chi}_yR||_{{\cal B}_p}\sum_{k=N+1}^{\infty} ||V_zR||_{\cal B}^k 
 \le 2^{-N}||{\chi}_yR||_{{\cal B}_p}.\eqno (3.28) 
 $$
Ainsi pour montrer (3.25(1)) il suffit de savoir que 
${\Phi}_{y,1,N}\in C_{\rm pp}({\R}^{d_1},{\cal B}_p)$. 
Mais les fonctions $z\to (V_zR)^k$ appartiennent \`a 
$C_{\rm pp}({\R}^{d_1},{\cal B})$ et ${\chi}_yR\in {\cal B}_p$,  
donc il est clair que 
${\Phi}_{y,1,N}\in C_{\rm pp}({\R}^{d_1},{\cal B}_p)$. 
\par 
Utilisant le raisonnement par r\'ecurrence on suppose que 
l'assertion (3.25($m$)) est vraie pour un certain $m\in {\N}^*$. 
Alors le Lemme 3.3 assure que   
 $$
 {\Phi}_{y,m+1,N}=\sum_{y'\in {\lp -N;\; N\lp}^d} {\Phi}_{y,1,N}
 {\Phi}_{y',m}\in C_{\rm pp}({\R}^{d_1},{\cal B}_{p/(m+1)}) 
 $$  
et utilisant Proposition 3.4 on peut estimer 
 $$
 N'>N\Rightarrow ||{\Phi}_{y,m+1,N'}(z)-{\Phi}_{y,m+1,N}(z)
 ||_{{\cal B}_{p/(m+1)}}\le
 $$
 $$
 \sum_{y'\in {\lp -N';\; N'\lp}^d\setminus {\lp -N;\; N\lp}^d} 
 ||{\chi}_yR_z{\chi}_{y'}||_{{\cal B}_p} 
 ||{\chi}_{y'}R^m_z||_{{\cal B}_{p/(m+1)}} 
 $$ 
 $$
 \le \sum_{y'\in {\Z}^d\setminus {\lp -N;\; N\lp}^d} 
 C(1+|y-y'|)^{-d-1} \, \le C'/N,
 $$
donc ${\Phi}_{y,m+1}$ est la limite de la suite 
$({\Phi}_{y,m+1,N})_{N\in \N}$ dans 
$C_{\rm pp}({\R}^{d_1},{\cal B}_{p/(m+1)})$. $\triangle$
 \par \bigskip 
\T 
\begin{prop} Soient $f\in C_0^{\infty}(\R )$, $L'\in \N$, $\rho >0$ 
et $y_1\in {\Z}^{d_1}$. On pose 
 $$
 u^{h,L'}_{\rho ,y_1}(z)={\rm tr}^h\, {\chi}^{h,L'}_{\rho ,y_1} 
(f(H^h_z)-f(-{\D}^h )), \eqno (3.29)
 $$
o\`u ${\chi}^{h,L'}_{\rho ,y_1}\in {\cal B}^h$ est d\'efini 
par la formule
 $$
 ({\chi}^{h,L'}_{\rho ,y_1}\f )(hn_1,hn_2)=
 \sum_{y_2\in {\lp -L';\; L'\lp}^{d_2} } 
 {\chi}_{{\cal C}(y_1,y_2)} ({hn_1/\rho},hn_2) 
 \f (hn_1,hn_2) \eqno (3.30)
 $$
pour $(n_1,n_2)\in {\Z}^{d_1}\times {\Z}^{d_2}$ et 
$\f \in l^2(h{\Z}^d)$. \par 
Si $\e >0$, $\rho \in [{1\over 2};\, 1]$ 
alors on peut trouver $N(\e )\in \N$, $c^{h,\rho}_{\e ,k}\in \C$, 
${\g}_{\e ,k}\in {\R}^{d_1}$  ($k=0,\dots ,N(\e )$) tels que    
 $$
 \sup_{0<h\le 1} \sup_{z\in {\R}^{d_1}} \left| {u^{h,L'}_{\rho ,0}
 (z)-\sum_{k=0}^{N(\e )} c^{h,\rho}_{\e ,k}{\rm e}^{iz{\g}_{\e ,k}} 
 }\right| <\e /8 , \eqno (3.31)
 $$ 
 $$
 \sup_{0<h\le 1} |c^{h,\rho}_{\e ,k}|<\infty . \eqno (3.31')
 $$
 \end{prop} 
{\underline {\it Preuve du fait que Proposition $3.5$ implique 
Th\'eor\`eme $1.2$}} 
\par \smallskip 
\T Au d\'ebut on observe que pour $\rho \in h\N$,$y_1\in {\Z}^{d_1}$ 
on peut \'ecrire 
 $$
 u^{h,L'}_{\rho ,0}(\rho y_1)={\rm tr}^h\, T^h_{(\rho y_1,0)} 
 {\chi}^{h,L'}_{\rho, y_1} (f(H^h_{\rho y_1})-f(-{\D}^h ))
 T^h_{(-\rho y_1,0)}, \eqno (3.32)
 $$
o\`u $T^h_{(\rho y_1,0)}$ est l'op\'erateur de la translation, 
 $$
 (T^h_{(\rho y_1,0)}\f )(hn)=\f (hn_1-\rho y_1,hn_2)\T 
 \hbox{ pour } \f \in l^2(h{\Z}^d).\eqno (3.33)
 $$
Compte tenu de 
 $$
 T^h_{( \rho y_1,0)}{\chi}^{h,L'}_{\rho ,0} 
 T^h_{(-\rho y_1,0)}={\chi}^{h,L'}_{\rho ,y_1},
 \hspace{7mm}
 T^h_{(\rho y_1,0)}H^h_{\rho y_1}T^h_{(-\rho y_1,0)}=H^h\eqno (3.34)
 $$
on obtient
 $$
 u^{h,L'}_{\rho ,0}(\rho y_1)={\rm tr}^h\, 
 {\chi}^{h,L'}_{\rho ,y_1} (f(H^h_z)-f(-{\D}^h)).\eqno (3.35)
 $$
En introduisant la notation $[L]_{\rho}=\rho [L/\rho ]$ on trouve 
 $$
 \sum_{y_1\in {\lp -L/\rho ;\; L/\rho \lp}^{d_1} } 
 {\chi}^{h,L'}_{\rho ,y_1} \, ={\chi}^{h,L'}_{[L]_{\rho}}\eqno (3.36)
 $$
et par cons\'equent 
 $$
 (2[L]_{\rho})^{-d_1}\sum_{y_1\in {\lp -L/\rho ;\; L/\rho \lp}^{d_1}} 
  u^{h,L'}_{\rho ,0}(\rho y_1)\, =N^{L'}_{[L]_{\rho}}(f,H^h).
\eqno (3.37)
 $$
Compte tenu du Corollaire 2.2 il suffit de montrer que pour tout 
$\e >0$ il existe $L_{\e}>0$ tel que 
 $$
 L_2\ge L_1\ge L_{\e}\Rightarrow 
 |N^{L'}_{L_1}(f,H^h)-N^{L'}_{L_2}(f,H^h)|<\e .\eqno (3.38)
 $$
Choisissant $L_{\e}$ assez grand et utilisant Corollaire 2.2(c) 
on voit que pour obtenir (3.38) il suffit de montrer 
que pour un choix convenable de 
$\rho (h,\e )\in [1/2;\; 1]\cap h\N$ on ait  
 $$
 L_2\ge L_1\ge L_{\e}\Rightarrow 
 |N^{L'}_{[L_1]_{\rho (h,\e )}}(f,H^h)-
 N^{L'}_{[L_2]_{\rho (h,\e )}}(f,H^h)|<\e /2. \eqno (3.39)
 $$ 
Soient $c^{h,\rho }_{\e ,k}$, ${\g}_{\e ,k}$ choisis comme dans la 
Proposition 3.5 et 
 $$
 N^{h,\rho }_{\e ,L}=\sum_{k=0}^{N(\e )} 
 c^{h,\rho }_{\e ,k} N_L^{\rho }({\g}_{\e ,k}),\eqno (3.40)
 $$
o\`u 
 $$
 N^{\rho }_L({\g}_{\e ,k})= (2[L]_{\rho })^{-d_1} 
 \sum_{y_1\in {\lp -L/\rho ;\; L/rho \lp}^{d_1}} 
 {\rm e}^{i\rho  y_1{\g}_{\e ,k}}.\eqno (3.40')
 $$
Alors (3.31) avec $\rho =\rho (h,\e )$ assure
 $$
 |N^{L'}_{[L]_{\rho (h,\e )}}(f,H^h)-
 N^{h,\rho (h,\e )}_{\e ,L}|<\e /8 \eqno (3.41)
 $$ 
et pour obtenir (3.39) il suffit de montrer 
 $$
 L_2\ge L_1\ge L_{\e}\Rightarrow 
 |N^{h,\rho (h,\e )}_{\e ,L_1}-
 N^{h,\rho (h,\e )}_{\e ,L_2}|<\e /4. \eqno (3.42)
 $$ 

Ainsi il suffit de prouver que $(3.40')$ poss\`ede une limite quand  
${L\to \infty}$ et le fait que pour tout $\e >0$ on peut trouver 
$h_{\e}>0$, $C_{\e}$ tels que 
 $$
 \sup_{0<h\le h_{\e}}\left| { N_L^{\rho (h,\e )}({\g}_{\e ,k})-
 \lim_{L\to \infty} N_L^{\rho (h,\e )}({\g}_{\e ,k}) }\right| 
 \le C_{\e}/L.\eqno (3.43(k)).
 $$
On peut supposer ${\g}_{\e ,0}=0$ (avec la possibilit\'e d'avoir 
$c^h_{\e ,0}=0$) et ${\g}_{\e ,k}\ne 0$ pour $k=1,...,N(\e )$. 
Alors $(3.43(0))$ est \'evident et  
pour d\'emontrer $(3.43(k))$ avec $k\ge 1$ on pose  
 $$
 d_{\e}(t)=\inf_{1\le k\le N(\e )} {\rm dist}
 (t{\g}_{\e ,k}, 2\pi {\Z}^d)\T \hbox{ pour } t>0.\eqno (3.44)
 $$
 Alors la fonction $t\to d_{\e}(t)$ est continue et 
$\{ t\in \R : d_{\e}(t)=0\}$ est discret. Ainsi  
on peut trouver $t(\e )\in [1/2;\; 3/4]$ tel que $d_{\e}(t)>0$ 
et $h(\e )>0$ tel que 
$t(\e )\le t\le t(\e )+h(\e )\Rightarrow d_{\e}(t(\e ))/2$,  
donc il existe $\rho (h,\e )\in [1/2;\; 1]\cap h\N$ tel que 
$d_{\e}(\rho (h,\e ))\ge d_{\e}(t(\e ))/2$.
Pareillement comme au d\'ebut de cette section on peut 
estimer    
 $$
 |N_L^{\rho (h,\e )}({\g}_{\e ,k})| \le 
 2L^{-1}|{\rm e}^{id_{\e}(t(\e ))/2}-1|^{-1}.\eqno (3.45) 
 $$ 
Ainsi on a d\'emontr\'e que Th\'eor\`eme 1.2 r\'esulte de 
Proposition 3.5.  $\triangle$ 
\par \bigskip 
{\it Preuve de la Proposition $3.5$} \T Soit 
${\cal A}=({\cal A}^h)_{0<h\le 1}$ une famille d'espaces 
de Banach ${\cal A}^h$. Alors on \'ecrira  
$\Phi \in C_{\rm pp}({\R}^{d_1};\; {\cal A})$ 
si et seulement si $\Phi =({\Phi}^h)_{0<h\le 1}$ est une famille 
d'applications ${\Phi}^h:\, {\R}^{d_1}\to {\cal A}^h$
telle  que  pour tout $\e >0$ on peut 
trouver $N(\e )\in \N$, ${\g}_{\e ,k}\in {\R}^{d_1}$, 
$A^h_{\e ,k}\in {\cal A}^h$ $(k=0,...,N(\e ))$ tels que 
 $$
 \sup_{0<h\le 1} \sup_{z\in {\R}^{d_1}}||{\Phi}^h(z)
 -{\sum}_{j=0}^{N(\e )} A^h_{\e ,k}\, {\rm e}^{iz\cdot {\g}_{\e ,k}}
  ||_{{\cal A}^h}\, <\e  \eqno (3.46)
 $$  
 et $\sup_{0<h\le 1} ||A^h_{\e ,k}||_{{\cal A}^h}<\infty$ pour 
$k=0,...,N(\e )$. \par 
Alors il est clair que $(z\to V^h_z)_{0<h\le 1}\in 
C_{\rm pp}({\R}^{d_1};\; ({\cal B}^h)_{0<h\le 1})$ et 
le Lemme 3.3 reste valable \'egalement.  
  
Il reste \`a suivre la preuve de la Proposition 3.2 
en rempla{\c{c}}ant $R_z$, $V_z$, $R$, ${\cal B}_{p/m}$ par 
$(R_z^h)_{0<h\le 1}$, $(V^h_z)_{0<h\le 1}$, 
$((-{\D}^h+{\la}_0I)^{-1})_{0<h\le 1}$ et 
$({\cal B}^h_{p/m})_{0<h\le 1}$ respectivement. $\triangle$

\section{Preuve du Th\'eor\`eme 1.5} 

 On note par ${\cal F}$ l'op\'erateur unitaire sur $L^2({\R}^d)$ 
donn\'e par la formule 
 $$
 ({\cal F}\f )(\xi )=(2\pi )^{-d/2}
 \int_{{\R}^d} {\rm e}^{ix\cdot \xi} \f (x)\, dx 
\T \hbox{ pour } \f \in L^1\cap L^2({\R}^d)  \eqno (4.1)
 $$ 
et on introduit la bijection isom\'etrique 
${\cal F}^h:\, l^2(h{\Z}^d)\to L^2({[-\pi /h;\; \pi /h[^d})$ 
par la formule 
 $$
 ({\cal F}^h\f )(\xi )=(h/2\pi )^{d/2}
 \sum_{n\in {\Z}^d} {\rm e}^{ihn\cdot \xi} \f (hn) 
\T \hbox{ pour } \f \in l^1\cap l^2(h{\Z}^d).\eqno (4.2)
 $$

\begin{lem} Soit $\th \in C_0^{\infty}({\R}^d)$. On d\'efinit  
${\th}^h\in l^2(h{\Z}^d)$ en posant ${\th}^h(hn)={\th} (hn)$ 
pour $n\in {\Z}^d$. Alors pour tout $N\in \N$ on a 
 $$
 \sup_{0<h\le 1} \T \sup_{\xi \in [-\pi /h;\; \pi /h[^d} h^{d/2}
 |\xi |^N\, |({\cal F}^h{\th}_h)(\xi )|\, < \infty .  \eqno (4.3)
 $$
\end{lem} 
{\it Preuve.} Pour $N\in \N$ posons  ${\th}_N^h=(-{\D}^h)^N{\th}^h$. 
Il est clair que 
 $$
 \sup_{n\in {\Z}^d}|{\th}_1^h(hn)|\le C\sup_{x\in {\R}^d} 
 \max_{1\le j\le d} |{\p}^2_{x_j}{\th}(x)| \le C'\eqno (4.4)
 $$
et plus g\'en\'eralement pour tout $N\in \N$ on a
 $$
 C_N=\sup_{0<h\le 1}\sup_{n\in {\Z}^d}|{\th}^h_N(hn)|<\infty ,
 \eqno (4.4')
 $$
donc  
 $$
 |({\cal F}^h{\th}^h_N)(\xi )|\le (h/2\pi )^{d/2}
 \sum_{\{ n\in {\Z}^d:\, hn\in {\rm supp\,}{\th}^h_N\} }
 |{\th}^h_N(hn)|\, \le C_Nh^{-d/2}.\eqno (4.5)
 $$
De l'autre c\^ot\'e on a $({\cal F}^h{\th}^h_N)(\xi )=
{\vt}_h(\xi )^N({\cal F}^h{\th}^h)(\xi )$ avec 
 $$
 {\vt}_h (\xi )=\sum_{j=1}^d {2-2\cos (h{\xi}_j)\over h^2}.
 \eqno (4.6)
 $$ 
Il reste \`a remarquer que     
 $$
 \xi \in [-\pi /h;\; \pi /h[^d\T \Longrightarrow \T \hbox{$1\over 4$} 
 |\xi |^2\,   \le {\vt}_h (\xi )\, \le |\xi |^2  \eqno (4.7)
 $$
avec (4.5) impliquent (4.3). $\triangle$

 \par \bigskip 

On note par $J^h$ l'injection isom\'etrique 
$L^2([\pi /h;\; \pi /h[^d)\to L^2({\R}^d)$, qui s'obtient 
en prolongeant les fonctions par la valeur 0 sur 
${\R}^d\setminus [-\pi /h;\; \pi /h[^d$. Alors   
 $$
 {J^h}^*\f ={\f}_{|_{[-\pi /h;\; \pi /h[^d}}\in  
 L^2([-\pi /h;\; \pi /h[^d) \T \hbox{ pour } \f \in L^2({\R}^d)
 \eqno (4.8)
 $$
 est la formule de l'op\'erateur adjoint \`a $J^h$. \par 
Pour $T>0$ on d\'efinit ${\chi}_T\in {\cal B}$ par la formule
 $$
 ({\chi}_T\f )(\xi )={\chi}_{[-T;\; T[^d}(\xi )\f (\xi )\T 
 \hbox{ pour } \f \in L^2({\R}^d). \eqno (4.9)
 $$  
Ensuite on d\'efinit $P_T\in {\cal B}$ et $P^h_T\in {\cal B}^h$ par  
 $$
 P_T={\cal F}^* {\chi}_T {\cal F},\hspace{9mm} 
  P^h_T=(J^h{\cal F}^h)^* {\chi}_T J^h {\cal F}^h  \eqno (4.10)
 $$  
et on remarque que $T\ge \pi /h\Rightarrow P^h_T=I$.

\begin{lem} Si $T'\ge 1$ est fix\'e, alors 
 $$
 \lim_{T\to \infty} || (I-P_T)\, \Th \, P_{T'}||_{\cal B}\, =0,  
 \eqno (4.11) 
 $$ 
 $$
 \lim_{T\to \infty} \sup_{0<h\le 1} || (I-P^h_T)\, {\Th}^h \, 
 P^h_{T'}||_{{\cal B}^h}\, =0.  \eqno (4.11')   
 $$
\end{lem} 
{\it Preuve.} Soit ${\widetilde \Th}^h_{T,T'}\in 
{\cal B}(L^2({[-\pi /h;\; \pi /h[^d}))$ d\'efini par  
 $$
 {\widetilde \Th}^h_{T,T'}={\cal F}^h(I-P^h_T){\Th}^hP^h_{T'} 
 ({\cal F}^h)^*.\eqno (4.12)
 $$
 Alors 
 $$
 ({\widetilde \Th}^h_{T,T'}\f )(\xi )=\int_{[-\pi /h;\; \pi /h[^d} 
 K^h_{T,T'}(\xi ,\xi ')\f (\xi ')\, d\xi ',\eqno (4.12')
 $$
 $$
 K^h_{T,T'}(\xi ,\xi ')=(1-{\chi}_{[-T;\; T[^d}(\xi ))
 (h/2\pi )^{d/2}({\cal F}^h{\th}^h)(\xi -\xi ')
 {\chi}_{[-T';\; T'[^d}(\xi '). \eqno (4.12'')
 $$
En vertu du Lemme 4.1 pour $T>T'$ on peut estimer  
 $$
 |K^h_{T,T'}(\xi ,\xi ')|\le (1-{\chi}_{[-T;\; T[^d}(\xi ))
 C_N(1+|\xi -\xi '|)^{-N}{\chi}_{[-T';\; T'[^d}(\xi ')
 $$
 $$
 \le C_N|T-T'|^{-N/2}(1+|\xi -\xi '|)^{-N/2}
 {\chi}_{[-T';\; T'[^d}(\xi ').\eqno (4.13)
 $$
Si $N>d$ alors on peut estimer
 $$
 ||{\widetilde \Th}^h_{T,T'}
 ||^2_{{\cal B}_2(L^2({[-\pi /h;\; \pi /h[^d}))}=
 \int_{{\R}^{2d}} |K^h_{T,T'}(\xi ,\xi ')|^2\, d\xi d\xi '\, \le 
 $$
 $$
 \int_{[-T';\; T'[^d} d\xi '\int_{[-\pi /h;\; \pi /h[^d} d\xi 
 C^2_N|T-T'|^{-N}(1+|\xi -\xi '|)^{-N}\, \le C'_N|T-T'|^{-N}
 $$
et pour terminer la preuve de (4.11$'$) on remarque 
 que pour $T'$ fix\'e on a 
 $$
 ||(I-P_T){\Th}^h P_{T'}||_{{\cal B}^h}=
 ||{\widetilde \Th}^h_{T,T'}
 ||_{{\cal B}(L^2({[-\pi /h;\; \pi /h[^d}))}
 $$
 $$
 \le ||{\widetilde \Th}^h_{T,T'}
 ||_{{\cal B}_2(L^2({[-\pi /h;\; \pi /h[^d}))}\to 0\T 
 \hbox{ quand } T\to \infty .
 $$  
\par \smallskip 
La preuve de (4.11) est semblable: 
$\th \in C_0^{\infty}({\R}^d)$ assure le fait que 
 ${\cal F}{\th}$ est \`a d\'ecroissance rapide, c'est-\`a-dire 
pour tout $N\in \N$ il existe $C_N>0$ telle que 
 $$
 |{\cal F}{\th}(\xi )|\le C_N(1+|\xi |)^{-N} \eqno (4.14)
 $$ 
 et d\'efinissant l'op\'erateur   
${\widetilde \Th}_{T,T'}={\cal F} (I-P_T){\Th}P_{T'}{\cal F}^*$ 
 on trouve l'expression  
 $$
 ({\widetilde \Th}_{T,T'}\f )(\xi )=\int_{{\R}^d} 
 K_{T,T'}(\xi ,\xi ')\f (\xi ')\, d\xi ',\eqno (4.15)
 $$
 $$
 K_{T,T'}(\xi ,\xi ')=(1-{\chi}_{[-T;\; T[^d}(\xi ))
 (2\pi )^{-d/2}({\cal F}{\th})(\xi -\xi ')
 {\chi}_{[-T'\;\; T'[^d}(\xi '),   \eqno (4.15')
 $$
 permettant d'appliquer le m\^eme raisonnement qu'auparavant.   
 $\triangle$
\par 
\begin{lem} On suppose $\th \in C_0^{\infty}({\R}^d)$ et 
$\Th$, ${\Th}^h$ comme dans Corollaire $2.4$. On pose 
 $$
 {\widetilde \Th}={\cal F} \Th {\cal F}^*,\hspace{8mm}
 {\widetilde \Th}^h=J^h{\cal F}^h {\Th}^h (J^h{\cal F}^h)^*,
 \eqno (4.16)
 $$
 $$
 {\widetilde V}={\cal F} V {\cal F}^*,\hspace{8mm}
 {\widetilde V}^h=J_h{\cal F}^h V^h (J^h{\cal F}^h)^*.\eqno (4.17)
 $$
Alors pour tout ${T'}>0$ on a
 $$
 \lim_{h\to 0} ||({\widetilde \Th}-{\widetilde \Th}^h)
 {\chi}_{T'} ||_{\cal B}\, =0, \hspace{7mm}
 \lim_{h\to 0} ||({\widetilde V}-{\widetilde V}^h)
 {\chi}_{T'} ||_{\cal B}\, =0.\eqno (4.18) 
 $$
\end{lem} 
{\it Preuve.} Soit $\e >0$. On va montrer qu'il existe  
$h_{\e}>0$ tel que  
 $$
 \sup_{0<h\le h_{\e}} ||({\widetilde \Th}-{\widetilde \Th}^h)
 {\chi}_{T'} ||_{\cal B} <\e .\eqno (4.19)
 $$
Mais Lemme 4.2 assure qu'il existe $T=T(\e )\ge T'$ tel que 
 $$
 ||(I-{\chi}_T){\widetilde \Th}{\chi}_{T'} ||_{\cal B}= 
 ||(I-P_T)\Th P_{T'} ||_{\cal B}<\e /3, \eqno (4.20)
 $$
 $$
 ||(I-{\chi}_T){\widetilde \Th}^h{\chi}_{T'} ||_{\cal B}= 
 ||(I-P^h_T){\Th}^hP^h_{T'} ||_{{\cal B}^h}<\e /3,\eqno (4.20') 
 $$
donc pour obtenir (4.18) il suffit de prouver 
 $$
\sup_{0<h\le h_{\e}} ||{\chi}_{T}({\widetilde \Th}-{\widetilde \Th}^h)
 {\chi}_{T'} ||_{\cal B} <\e /3 \eqno (4.21)
 $$  
On peut supposer que $h\le h_{\e}\le \pi /T\le \pi T'$. Alors 
 $$
 ({\chi}_{T}({\widetilde \Th}^h-{\widetilde \Th})
 {\chi}_{T'})\f (\xi )=\int_{{\R}^d} 
 {\widetilde K}^h_{T,T'}(\xi ,\xi ')\f (\xi ')\, d\xi ',\eqno (4.22)
 $$ 
 $$
 {\widetilde K}^h_{T,T'}(\xi ,\xi ')=(2\pi )^{-d/2} 
 {\chi}_{[-T;\; T[^d}(\xi )(h^d{\cal F}^h{\th}^h-{\cal F}{\th})
 (\xi -\xi '){\chi}_{[-T';\; T'[^d}(\xi ').\eqno (4.22')
 $$
Ensuite on remarque que 
 $$
 \lim_{h\to 0} {\widetilde K}^h_{T,T'}(\xi ,\xi ')\, = 0
\hbox{ pour } \xi ,\xi '\in {\R}^d.\eqno (4.23)
 $$
En effet, pour tout $\xi \in {\R}^d$ on a 
 $h^d{\cal F}^h{\th}^h(\xi )\to {\cal F}{\th}(\xi )$ quand $h\to 0$, 
parce qu'il s'agit de la suite des sommes 
de Riemann de l'int\'egrale de la fonction appartenant \`a 
$C_0^{\infty}({\R}^d)$. \par 
Pour terminer la preuve de la premi\`ere assertion (4.18) 
on remarque qu'il existe $C_0>0$ telle que 
$|{\widetilde K}^h_{T,T'}(\xi ,\xi ')|\le C_0{\chi}_{[-T;\; T[^{2d}}
 (\xi ,\xi ')$, alors le th\'eor\`eme de la 
convergence domin\'ee de Lebesgue permet de conclure que (4.23) 
implique  
 $$
 \limsup_{h\to 0} ||{\chi}_{T}({\widetilde \Th}-{\widetilde \Th}^h) 
 {\chi}_{T'} ||^2_{\cal B}\le \limsup_{h\to 0} ||{\chi}_{T}( 
 {\widetilde \Th}-{\widetilde \Th}^h){\chi}_{T'}||^2_{{\cal B}_2} 
 $$
 $$
 \le \lim_{h\to 0} \int_{{\R}^{2d}} 
 |{\widetilde K}^h_{T,T'}(\xi ,\xi ')|^2\, d\xi \, d\xi '\, = 0.
 $$ 
\par \bigskip 
Pour terminer la preuve de la deuxi\`eme assertion (4.18) 
on doit trouver $h_{\e}>0$ tel que 
 $$
 \sup_{0<h\le h_{\e}} ||({\widetilde V}-{\widetilde V}^h)
 {\chi}_{T'} ||_{\cal B} <\e .\eqno (4.24)
 $$
Le Lemme 2.4 assure l'existence de $N(\e )\in \N$, 
${\g}_{\e ,k}\in {\R}^{d_1}$, $v_{\e ,k}\in C_0^{\infty}({\R}^{d_1})$, 
$k=1,...,N(\e )$ tels que l'on ait $||V-V_{\e}||_{\cal B}<\e /3$ 
avec
 $$
 V_{\e}=\sum_{k=1}^{N(\e )} e_{{\g}_{\e ,k}}\otimes V_{\e ,k},
 $$
o\`u $e_{{\g}_{\e ,k}}\in {\cal B}(L^2({\R}^{d_1}))$, 
$V_{\e ,k}\in {\cal B}(L^2({\R}^{d_2}))$ sont donn\'es par 
 $$
 (e_{{\g}_{\e ,k}}{\f}_1)(x_1)={\rm e}^{ix_1{\g}_{\e ,k}}{\f}_1(x_1)
 \T \hbox{ pour } \T {\f}_1\in L^2({\R}^{d_1}),
 $$ 
 $$
 (V_{\e ,k}{\f}_2)(x_2)=v_{\e ,k}(x_2){\f}_2(x_2)
 \T \hbox{ pour } \T {\f}_2\in L^2({\R}^{d_2}).
 $$
Alors on a \'egalement $||V^h-V^h_{\e}||_{{\cal B}^h}<\e /3$ avec 
 $$
 V^h_{\e}=\sum_{k=1}^{N(\e )} e^h_{{\g}_{\e ,k}}\otimes V^h_{\e ,k},
 $$
o\`u $e^h_{{\g}_{\e ,k}}\in {\cal B}(l^2(h{\Z}^{d_1}))$, 
$V^h_{\e ,k}\in {\cal B}(l^2(h{\Z}^{d_2}))$ sont donn\'es par 
 $$
 (e^h_{{\g}_{\e ,k}}{\f}_1)(hn_1)={\rm e}^{ihn_1{\g}_{\e ,k}}{\f}_1   (hn_1)\T \hbox{ pour } \T {\f}_1\in l^2(h{\Z}^{d_1}),
 $$ 
 $$
 (V^h_{\e ,k}{\f}_2)(hn_2)=v_{\e ,k}(hn_2){\f}_2(hn_2)
 \T \hbox{ pour } \T {\f}_2\in h^2(h{\Z}^{d_2}).
 $$
Ainsi au lieu de (4.24) il suffit de prouver 
 $$
 \sup_{0<h\le h_{\e}} ||({\widetilde V}_{}-{\widetilde V}^h)
 {\chi}_{T'} ||_{\cal B} <\e /3  \eqno (4.25)
 $$
avec  
 $$ 
 {\widetilde V}_{\e}={\cal F}V_{\e}{\cal F}^*,\hspace{1cm} 
 {\widetilde V}^h_{\e}=J^h{\cal F}^hV^h_{\e}(J^h{\cal F}^h)^*.
 $$ 
 Ensuite on remarque que ${\cal F}={\cal F}_1\otimes {\cal F}_2$, 
o\`u  ${\cal F}_j$  pour $j=1$, 2, est l'op\'erateur unitaire sur 
 $L^2({\R}^{d_j})$ donn\'e par la formule 
 $$
 ({\cal F}_j{\f}_j)({\xi}_j)=(2\pi )^{-d_j/2}
 \int_{{\R}^{d_j}} {\rm e}^{ix_j\cdot {\xi}_j} {\f}_j(x_j)\, dx_j 
\T \hbox{ pour } {\f}_j\in L^2({\R}^{d_j})  
 $$ 
et introduisant 
${\widetilde V}_{\e ,k}={\cal F}_2V_{\e ,k}{\cal F}_2^*$ on trouve 
 $$
 {\widetilde V}_{\e}=
 \sum_{k=1}^{N(\e )} T_{{\g}_{\e ,k}}\otimes {\widetilde V}^h_{\e ,k}
 $$ 
o\`u $(T_{{\g}_{\e ,k}}{\f}_2)(x_2)={\f}_2(x_2-{\g}_{\e ,k})$ 
pour ${\f}_2\in L^2({\R}^{d_2})$.
 \par 
 De mani\`ere analogue 
${\cal F}^h={\cal F}^h_1\otimes {\cal F}^h_2$ avec  
${\cal F}_j^h:\, l^2(h{\Z}^{d_j})\to L^2([-\pi /h;\; \pi /h[^{d_j})$ 
pour $j=1$, 2, est donn\'e par la formule 
 $$
 ({\cal F}_j^h{\f}_j)({\xi}_j)=(h/2\pi )^{d_j/2}
 \sum_{n\in {\Z}^{d_j}} {\rm e}^{ih{n_j}\cdot {\xi}_j}{\f}_j(hn_j) 
 \T \hbox{ pour } {f_j}\in l^1\cap l^2(h{\Z}^{d_j}) 
 $$
et introduisant 
${\widetilde V}^h_{\e ,k}=
 J^h{\cal F}^h_2V^h_{\e ,k}(J^h{\cal F}^h_2)^*$ on trouve que 
pour $h\le \pi /T$ on a
 $$
 {\widetilde V}^h_{\e}{\chi}_T=\sum_{k=1}^{N(\e )} T_{{\g}_{\e ,k}}   {\chi}^{(1)}_T \otimes {\widetilde V}^h_{\e ,k}{\chi}^{(2)}_T
 $$ 
o\`u ${\chi}^{(j)}_T\in {\cal B}(L^2({\R}^{d_j})$ avec $j=1$, 2, 
est donn\'e par la formule 
 $$ 
 ({\chi}^{(j)}_T{\f}_j)(x_j)={\chi}_{[-T;\; T[^{d_j}}(x_j)
 {\f}_j(x_j)\T \hbox{ pour } {\f}_j\in L^2({\R}^{d_j}).
 $$
Cependant la d\'emonstration de la partie (a) donne \'egalement
 $$
 \lim_{h\to 0} ||({\widetilde V}_{\e ,k}-{\widetilde V}^h_{\e ,k})
 {\chi}^{(2)}_T||_{{\cal B}(L^2({\R}^{d_2}))}\, =0 \eqno (4.26)
 $$
et compte tenu de
 $$
 ({\widetilde V}_{\e}-{\widetilde V}^h_{\e}){\chi}_T=
 \sum_{k=1}^{N(\e )} T_{{\g}_{\e ,k}} {\chi}^{(1)}_T \otimes 
({\widetilde V}_{\e ,k}-{\widetilde V}^h_{\e ,k}){\chi}^{(2)}_T
 $$
il est clair que (4.25) r\'esulte de (4.26). $\triangle$

\begin{lem} Soit ${\la}_0\ge 1+2||V||$ et posons  
 $$ 
 R=(-\D +V+{\la}_0I)^{-1}, \hspace{8mm} 
 R^h=(-{\D}^h+V^h+{\la}_0I)^{-1},\eqno (4.27)
 $$
 $$
 {\widetilde R}={\cal F} R {\cal F}^*, \hspace{9mm}
 {\widetilde R}^h=J^h{\cal F}^h V^h (J^h{\cal F})^*.\eqno (4.28)
 $$ 
Alors \T 
$\lim_{h\to 0} ||{\widetilde R}-{\widetilde R}^h||_{\cal B}\T =0$.
\end{lem} 
{\it Preuve.} Compte tenu de les expressions
 $$
 R={\widetilde R}_{\circ}\sum_{k=0}^{\infty} 
 (-{\widetilde V}{\widetilde R}_{\circ})^k,
 \hspace{8mm}
 R^h={\widetilde R}^h_{\circ}\sum_{k=0}^{\infty} 
 (-{\widetilde V}^h{\widetilde R}_{\circ}^h)^k,
 $$ 
o\`u on a d\'esign\'e  
 $$
 {\widetilde R}_{\circ}={\cal F}(-\D +{\la}_0I)^{-1}{\cal F}^*, 
 \hspace{8mm} {\widetilde R}^h_{\circ}=J^h{\cal F}^h
 (-{\D}^h+V^h+ {\la}_0I)^{-1}(J^h{\cal F})^*,
 $$ 
 il suffit de montrer 
 $$
 \lim_{h\to 0} ||{\widetilde R}-{\widetilde R}^h||_{\cal B}\T =0, 
 \hspace{8mm} \lim_{h\to 0} ||{\widetilde V}{\widetilde R}-
 {\widetilde V}^h{\widetilde R}^h||_{\cal B}\T =0.\eqno (4.29) 
 $$
D'abord 
 on montre que pour tout $\e >0$ il existe $h_{\e}>0$ tel que  
 $$
 \sup_{0<h\le h_{\e}} ||{\widetilde R}-{\widetilde R}^h||_{\cal B} 
 <\e .\eqno (4.30)
 $$
Cependant  
 $$
 ({\widetilde R}_{\circ}\f )(\xi )=(|\xi |^2+{\la}_0)^{-1}\f (\xi ),
 $$
 $$
 ({\widetilde R}^h_{\circ}\f )(\xi)={\chi}_{[-\pi /h;\; \pi /h[^d}
 (\xi )({\vt}_h(\xi)+{\la}_0)^{-1}\f (\xi ),
 $$
o\`u ${\vt}_h$ est donn\'e par (4.6) et on remarque qu'il existe 
$C_0>0$ tel que  
 $$
 ||(I-{\chi}_T){\widetilde R}_{\circ}||_{\cal B}=
 \sup_{\xi \in {\R}^d\setminus [-T;\; T[^d} (|\xi |^2+{\la}_0)^{-1}
 \, \le C_0T^{-2},\eqno (4.31)
 $$
 $$
 h\le \pi /T\, \Rightarrow \, ||(I-{\chi}_T){\widetilde R}_{\circ}^h
 ||_{\cal B}=\sup_{\xi \in {\R}^d\setminus [-T;\; T[^d} 
 ({\vt}_h(\xi)+{\la}_0)^{-1}\, \le C_0T^{-2}, \eqno (4.31')
 $$
o\`u la derni\`ere estimation r\'esulte de (4.7). Pour justifier 
(4.30) il suffit de choisir $T$ tel que $C_0T^{-2}<\e /4$ et 
utiliser le fait que 
 $$
 \lim_{h\to 0} \sup_{\xi \in [-T;\; T[^d} \left| {
 {\vt}_h(\xi)-|\xi |^2 }\right| \T =0.\eqno (4.32)
 $$   
Il reste \`a montrer l'existence de $h_{\e}>0$ tel que 
 $$
 \sup_{0<h\le h_{\e}} || {\widetilde V}^h{\widetilde R}_{\circ}-
 {\widetilde V} {\widetilde R}_{\circ}||_{\cal B} <\e .\eqno (4.33)
 $$
Soit $T$ tel que $2||V||_{\cal B}C_0T^{-2}<\e /4$. 
Si $h_{\e}>0$ est suffisamment petit, alors  
 $$
 \sup_{0<h\le h_{\e}} ||{\widetilde V}^h({\widetilde R}_{\circ}^h-
 {\widetilde R}_{\circ})||_{\cal B}\le ||V||_{\cal B}||
 {\widetilde R}_{\circ}^h-
 {\widetilde R}_{\circ})||_{\cal B}<\e /4\eqno (4.34)
 $$ 
 et compte tenu du Lemme 2.3,
 $$
 \sup_{0<h\le h_{\e}} ||({\widetilde V}^h-{\widetilde V}){\chi}_T
 ||_{\cal B} <\e /4.\eqno (4.35)
 $$
Pour obtenir (4.33) on estime 
 $$
 || {\widetilde V}^h{\widetilde R}_{\circ}-
 {\widetilde V} {\widetilde R}_{\circ}||_{\cal B}\le 
 ||{\widetilde V}^h({\widetilde R}_{\circ}^h-
 {\widetilde R}_{\circ})||_{\cal B} +
 $$
 $$
 ||({\widetilde V}^h-{\widetilde V}){\chi}_T
 ||_{\cal B} ||{\widetilde R}_{\circ}||_{\cal B}+
 ||{\widetilde V}^h-{\widetilde V}||_{\cal B} ||(I-{\chi}_T)
 {\widetilde R}_{\circ}||_{\cal B}
 $$
et (4.34), (4.35) permettent de majorer la derni\`ere expression 
par ${3\over 4}\e +2||V||_{\cal B}C_0T^{-2}<\e$. $\triangle$
 \par \bigskip 

\begin{lem} Pour d\'emontrer Th\'eor\`eme $1.5$ il suffit de montrer 
que l'on a 
$$
 \lim_{h\to 0} {\rm tr}^h\, P^h_T{\Th}^h(R^h)^m{\Th}^h\T =
\, {\rm tr}\, P_T\Th R^m{\Th}.\eqno (4.36)
 $$
pour tout $T\ge 1$ et $m\ge 2+{d\over 2}$. 
\end{lem} 
{\it Preuve.} 
\T Compte tenu du Corollaire 2.3 il faut montrer que pour 
tout $\e >0$ il est possible de trouver $h_{\e}>0$ tel que 
 $$
\sup_{0<h\le h_{\e}} \left| {{\rm tr}^h\, {\Th}^hf(H^h) {\Th}^h \, 
-\, {\rm tr}\, {\Th} f(H){\Th}}\right| <\e .\eqno (4.37)
 $$ 
D'abord on va montrer qu'il existe $T(\e )$ tel que pour 
$T\ge T(\e )$ on a 
 $$ 
  || (I-P^h_T){\Th}^hf(H^h){\Th}^h
 ||_{{\cal B}^h_1}<\e /4. \eqno (4.38) 
 $$
En effet, le membre gauche de (4.38) est major\'e par 
${\zeta}_1(h)+{\zeta}_2(h)$ avec 
  $$
 {\zeta}_1(h)=||(I-P^h_T){\Th}^hP^h_{T'}||_{{\cal B}^h}\, 
 ||f(H^h){\Th}^h||_{{\cal B}^h_1}, 
 $$
 $$
 {\zeta}_2(h)=||{\Th}^h(I-P^h_{T'})({\la}_0I-{\D}^h)^{-1}
 ||_{{\cal B}^h}\, ||({\la}_0I-{\D}^h)R^h||_{{\cal B}^h}\, 
 ||g(H^h){\Th}^h||_{{\cal B}^h_1} 
 $$
 o\`u on a not\'e $g(\la )=({\la}_0+\la )f(\la )$. Puisque 
  $$
 ||(I-P^h_{T'})({\la}_0I-{\D}^h)^{-1}||_{{\cal B}^h}\, =
 \sup_{\xi \in [-\pi /h;\; \pi /h[^d\setminus [-T';\; T'[^d} 
 |{\la}_0+{\vt}_h(\xi )|^{-1} \, \le CT'^{-2},
 $$
on peut choisir $T'$ suffisamment grand pour assurer 
${\zeta}_2(h)<\e /8$ pour tout $h\in ]0;\; 1]$ et ensuite Lemme 4.2  
permet de choisir $T$ suffisamment grand pour assurer 
${\zeta}_1(h)<\e /8$ pour tout $h\in ]0;\; 1]$. \par 
De mani\`ere analogue il est possible de trouver 
$T(\e )$ tel que   
 $$
 T\ge T(\e )\Rightarrow 
 ||(I-P_T){\Th}f(H){\Th}||_{{\cal B}_1}< \e /4.\eqno (4.39) 
 $$ 
Dans le deuxi\`eme pas on consid\`ere une approximation de $f$ par
 $(f_{\e '})_{\e '>0}$ comme dans la  d\'emonstration de 
Proposition 3.1. Alors il existe $C_0>0$ 
telle que pour tout $\e '>0$,  
 $$
 ||P^h_T{\Th}^h(f-f_{\e '})(H^h){\Th}^h||_{{\cal B}^h_1} \le 
 \e '\, ||(R^h)^m{\Th}^h||_{{\cal B}^h_1} \, \le C_0\e ', 
 $$
 $$
 \left| {{\rm tr}\, P_T{\Th}(f-f_{\e '})(H){\Th}}\right| \le 
 \e '\, ||(R^h)^m{\Th}||_{{\cal B}_1} \, \le C_0\e '. 
 $$
Soit $\e '=\e /(8C_0)$. Alors (4.37) r\'esulte de  
 $$
 \sup_{0<h\le h_{\e}} \left| {{\rm tr}^h\, P^h_T{\Th}^hf_{\e'}(H^h)
  {\Th}^h\, -\, {\rm tr}\, P_T{\Th}f_{\e'}(H){\Th}}\right| <\e /4  
   \eqno (4.40)
 $$
et on termine la d\'emonstration en observant que (4.36) assure 
l'existence de $h_{\e}>0$ tel que (4.40) soit satisfaite. $\triangle$   
 \par \bigskip 

{\underline {\it Preuve du Th\'eor\`eme $1.5$}}
 \par \smallskip 
La d\'emonstration est bas\'ee sur les \'egalit\'es
 $$
{\rm tr\,}P_T{\Th}R^m{\Th}P_T\, ={\rm tr}\, {\chi}_T{\widetilde \Th}   {\widetilde R}^m{\widetilde \Th}{\chi}_T,
 $$
 $$
 {\rm tr}^h\, P^h_T{\Th}^h(R^h)^m{\Th}^hP^h_T\, = {\rm tr}\, {\chi}_T  
 {\widetilde \Th}^h({\widetilde R}^h)^m{\widetilde \Th}^h{\chi}_T,
 $$
 qui permettent d'\'ecrire (4.36) sous la forme
 $$
 \lim_{h\to 0} {\rm tr}\, {\chi}_T\left( { {\widetilde \Th}^h 
 ({\widetilde R}^h)^m{\widetilde \Th}^h\, -\, {\widetilde \Th} 
 {\widetilde R}^m{\widetilde \Th}}\right){\chi}_T \, =\, 0.
 \eqno (4.41)
 $$
Pour d\'emontrer (4.41) on remarque que  
 $$
 ||{\chi}_T({\widetilde \Th}^h({\widetilde R}^h)^m{\widetilde \Th}^h
 \, -\, {\widetilde \Th}{\widetilde R}^m{\widetilde\Th}){\chi}_T 
 ||_{{\cal B}_1}\le {\zeta}_1(h)+{\zeta}_2(h)+{\zeta}_3(h) 
 $$
 avec
 $$ 
 {\zeta}_1(h)=||{\widetilde \Th}^h({\widetilde R}^h)^m||_{{\cal B}_1} 
 \, ||({\widetilde \Th}^h-{\widetilde \Th}){\chi}_T||_{\cal B},
 $$
 $$
 {\zeta}_2(h)=||{\widetilde \Th}^h(({\widetilde R}^h)^m-
 {\widetilde R}^m){\widetilde \Th}||_{{\cal B}_1}, 
 $$ 
 $$
 {\zeta}_3(h)=||{\chi}_T({\widetilde \Th}^h-{\widetilde \Th})
 ||_{\cal B}\, ||{\widetilde R}^m{\widetilde \Th}||_{{\cal B}_1}.
 $$
 En utilisant 
 $$
 ({\widetilde R}^h)^m-{\widetilde R}^m=\sum_{m'=1}^m 
 ({\widetilde R}^h)^{m'-1}({\widetilde R}^h-{\widetilde R})
 {\widetilde R}^{m-m'}
 $$
on peut estimer 
 $$
 {\zeta}_2(h)\le \sum_{m'=1}^m ||{\widetilde \Th}^h
 ({\widetilde R}^h)^{m'-1}||_{{\cal B}_{(m-1)/(m'-1)}}\, 
 ||{\widetilde R}^h-{\widetilde R}||_{\cal B}\, 
||{\widetilde R}^{m-m'}
 {\widetilde \Th}||_{{\cal B}_{(m-1)/(m-m')}} 
 $$
avec la convention que l'on utilise la norme $||\cdot ||_{\cal B}$ au 
lieu de $||\cdot ||_{{\cal B}_{(m-1)/(m'-1)}}$ dans le cas $m'=1$ et  au lieu de $||\cdot ||_{{\cal B}_{(m-1)/(m-m')}}$ dans le cas $m'=m$.
Compte tenu du Lemme 2.2 du Chapitre 3 on a
 $$
 \sup_{0<h\le 1} ||{\widetilde \Th}^h({\widetilde R}^h)^{m'-1}
 ||_{{\cal B}_{(m-1)/(m'-1)}}=\sup_{0<h\le 1} 
 ||{\Th}^h(R^h)^{m'-1}||_{{\cal B}_{(m-1)/(m'-1)}} <\infty ,
 $$ 
et de mani\`ere analogue on a
 $$
 ||{\widetilde R}^{m-m'}{\widetilde \Th}
 ||_{{\cal B}_{(m-1)/(m-m')}}=||R^{m-m'}\Th 
 ||_{{\cal B}_{(m-1)/(m-m')}}<\infty .
 $$ 
Ainsi utilisant le Lemme 4.5 
on trouve $\lim_{h\to 0} {\zeta}_2(h)\, =0$. 
Pour terminer la d\'emonstration on remarque  
de mani\`ere analogue que 
$\lim_{h\to 0} {\zeta}_1(h)\, =\lim_{h\to 0} {\zeta}_3(h)\, =0$ 
r\'esulte de Lemme 4.2.    
$\triangle$   
   
\end{document}